\documentclass[journal]{IEEEtran}
\usepackage{cite}
\usepackage{times,amsmath,amssymb,psfrag,stfloats}
\usepackage[demo]{graphicx}
\usepackage{epsfig}
\usepackage{color}
\usepackage{hyperref}
\usepackage{breakurl}
\usepackage{algpseudocode}
\usepackage{algorithm}
\usepackage{hyperref}
\usepackage{multirow}
\usepackage{epstopdf}
\usepackage{tabu}
\usepackage{float}
\usepackage{color}
\usepackage{nomencl}
\usepackage{rotating}

\usepackage{bm}
\usepackage{cases}
\usepackage{colortbl}
\usepackage{xcolor}
\usepackage{MnSymbol}
\usepackage{underscore}
\usepackage[font = small]{caption}

\newcommand{\ct}{\color{black}}

\newcommand{\cheewooi}{\color{black}}

\begin{document} \sloppy
\title{Impact Assessment of Hypothesized {\ct Cyberattacks} \\ on {\ct Interconnected Bulk Power Systems}}
\author{Chee-Wooi Ten,~\IEEEmembership{Senior Member,~IEEE}, Koji Yamashita,~\IEEEmembership{Member,~IEEE}, Zhiyuan Yang,~\IEEEmembership{Student Member,~IEEE}, \\Athanasios V. Vasilakos,~\IEEEmembership{Senior Member,~IEEE}, and Andrew Ginter,~\IEEEmembership{Member,~IEEE}
\thanks{This work was supported by the US National Science Foundation (NSF) on power grid resilience under the collaborative research awards (CNS\&CSE1541000 and ECCS1128512) titled ``Revolution through Evolution: A Controls Approach to Improve How Society Interacts with Electricity'' and ``Integrated Vulnerability-Reliability Modeling and Analysis of Cyber-Physical Power Systems,'' respectively.}
\thanks{C.-W. Ten, K. Yamashita, Z. Yang are with the Electrical and Computer Engineering Department, Michigan Technological University, Houghton, MI, 49931 USA (e-mails: ten@mtu.edu, kyamashi@mtu.edu, and yzhiyuan@mtu.edu).}
\thanks{A. V. Vasilakos is with Lule{\aa} University of Technology, Sweden (e-mail: athanasios.vasilakos@ltu.se).}
\thanks{A. Ginter is with the Waterfall Security Solutions, Isreal (e-mail: andrew.ginter@waterfall-security.com).}
}

\maketitle
\begin{abstract}
The first-ever Ukraine cyberattack on power grid has proven its devastation by hacking into their critical cyber assets. With administrative privileges accessing substation networks/local control centers, one intelligent way of coordinated cyberattacks is to execute a series of disruptive switching executions on multiple substations using compromised supervisory control and data acquisition (SCADA) systems. These actions can cause significant impacts to an interconnected power grid. Unlike the previous power blackouts, such high-impact initiating events can aggravate operating conditions, initiating instability that may lead to system-wide cascading failure. A systemic evaluation of ``nightmare'' scenarios is highly desirable for asset owners to manage and prioritize the maintenance and investment in protecting their cyberinfrastructure. This survey paper is a conceptual expansion of real-time monitoring, anomaly detection, impact analyses, and mitigation (RAIM) framework that emphasizes on the resulting impacts, both on steady-state and dynamic aspects of power system stability. Hypothetically, we associate the combinatorial analyses of steady state on substations/components outages and dynamics of the sequential switching orders as part of the permutation. The expanded framework includes (1) critical/non-critical combination verification, (2) cascade confirmation, and (3) combination re-evaluation. This paper ends with a discussion of the open issues for metrics and future design pertaining the impact quantification of cyber-related contingencies.
\end{abstract}

\begin{IEEEkeywords}
Consequential cyberattacks, cyber-physical security, power substations, {\cheewooi system resilience}, the I of RAIM.
\end{IEEEkeywords}
\IEEEpeerreviewmaketitle

\section{Development of Power Grid Cybersecurity}
\IEEEPARstart{T}{HE} smart grid roadmap has been envisioned to promote deploying advanced communication infrastructure as well as integrating distributed energy resources with existing power infrastructure. The digital upgrade of grid systems provides tremendous opportunities to integrate advanced sensing technologies, such as synchrophasor units across substations, for monitoring and control of a power grid. The new definition of grid reliability in this vision focuses on increased ability to withstand severe weather events, deliberate sabotage, and cyberattacks. Intelligent automation and communications infrastructure is intended to improve the accuracy of detection of abnormal conditions with ``self-healing'' capability. Such features have been implemented in power control center applications that facilitate monitoring and control. {\cheewooi Technological evolution of computerized management system on supervisory} control and data acquisition (SCADA) infrastructure can potentially be prone to electronic intrusions due to the vulnerabilities of access points in critical infrastructure \cite{5484019,6303885, 6307833}.

\begin{figure*}
\centering
\includegraphics[width = 18cm]{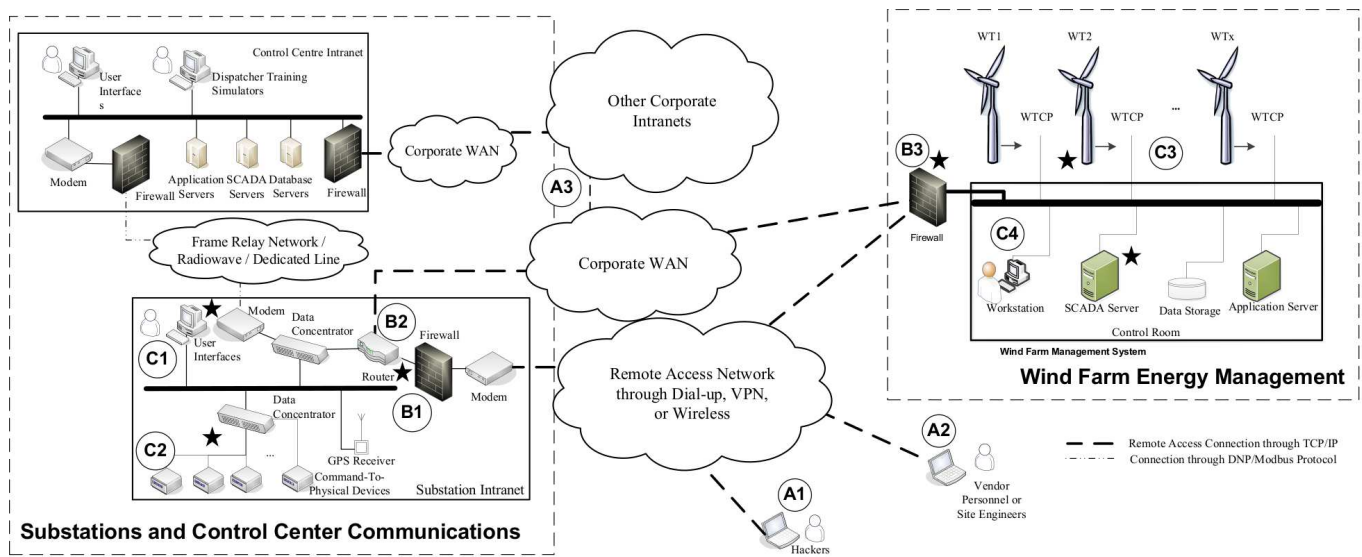}
  \caption{SCADA Infrastructure and Network Connectivity Between Substation-Level Network, Control Centers, and a Wind Farm Management System \cite{9000004,lingfeng2,lingfeng1}.}\label{SubControl} \vspace*{-5pt}
\end{figure*}
\subsection{Digital Modernization and Emerging Threats}
{\ct
There have been growing concerns over the potential of cyberattacks on the integration of IP-based communication infrastructure. Misconceptions of power grid cybersecurity have been reported \cite{21myth}. One real example of persistent threat is \emph{Stuxnet}. Comparable attack agents launched by a group of attackers or organizations able to systematically penetrate the generation control networks that would enable coordinated cyberattack in power grid SCADA networks have been hypothesized  \cite{2, 6112333}.  The sophistication of such persistent attacks will continue to evolve with domain-specific targets \cite{6112333}.  Another example is the 2011 \emph{Night Dragon} attack  \cite{3}. Clearly, coordinated attacks pose a great threat to the SCADA network of a nation’s power grid. A published workshop by the National Academies \cite{NAE} has indicated the trend of cyberattacks will further evolve towards more adaptive, coordinated, and persistent attack capabilities. Attackers can gain access to substation communication frameworks through access points such as gateways, leased-line and dial-up modems, virtual private networks (VPN), wireless links  \cite{9000004} as well as by physically breaking into unstaffed substations. Once attackers gain access to intelligent electronic devices (IEDs) they can access substation configuration descriptions and configurations in IED description files. By modifying the critical settings of specific IEDs and protective relays, intruder can disrupt the smooth operation of substation, in the worst case causing devastating failures of substation equipment and even cascading shutdown of other substations.

As depicted {\cheewooi generally} in Fig. \ref{SubControl}, the following {\cheewooi combinations enumerate} possible remote connections to local substation-level networks, describing intrusion paths from outside to the control networks.
\begin{itemize}
\setlength{\itemsep}{0pt}
\setlength{\parskip}{0pt}
\setlength{\parsep}{0pt}
  \item Any (A1,A2,A3)--B1--B2--C1
  \item Any (A1,A2,A3)--B1--B2--C2
  \item Any (A1,A2,A3)--B3--C3
\end{itemize}

Each combination includes connections through remote dial-up systems or {\cheewooi VPN} to substation-level networks targeted on substation user interface or {\cheewooi IEDs}. Once a local network has been compromised, a cyberattack can be launched either through (1) User interface, C1, (2) Direct IED connection, C2, or (3) Eavesdropping and data packet modification, C3. Once the network security has been compromised, there are two possibilities that allow the attackers to launch a cyber attack, i.e., (A) through user interface on the local network, or (B) direct access to IEDs (Both numbered as C1 and C2 in Fig. 1, respectively). Generally, the intrusion scenarios enable attackers to launch disruptive switching actions through (A) direct accessing substation IEDs  \cite{1IEC61850, 2IEC61850} or (B) accessing substation user interfaces in the local network. Due to the increasing penetration of renewable energy sources around the world, the curtailment of wind generation is required in some countries. This includes active power control and reactive power control within the plant-level control. The undesired tripping of the wind farm and deficiency of the secondary reserve as well as the supported reactive power could deteriorate voltage and frequency. An intelligent cyberattack to this control system, upon successful intrusion to the network, would be a new threat to overcome \cite{lingfeng2,lingfeng1}.}

\subsection{Technical Compliance of Federal Agencies}

The expanding IP-based communication infrastructure in power substations raises the likelihood of coordinated cyberattacks which threaten power grids with a large-scale blackout. Federal Energy Regulatory Commission’s (FERC) studies on United States power grids reveal that small-scale coordinated cyberattacks on multiple substations could cause a nationwide blackout \cite{fercstudy}. Over the past seven years, the Federal Energy Regulatory Commission (FERC) has mandated compliance with North American Electric Reliability Corporation (NERC) critical infrastructure protection (CIP) standards in ten mandatory areas including sabotage reporting and the implementation of physical and electronic security of cyber assets within electronic security perimeters. Although various versions of CIP standards have been amended to strengthen security protections, compliant sites are still at risk of malware-based attacks such as Stuxnet, able to propagate within substation-level wide-area Intranet networks. The consequences of potential cyberattacks can severely affect the reliability. A cyber-vulnerability assessment framework with improved component outage simulation could establish valuable new sets of operational and financial constraints for overall system planning. Hypothetical attack scenarios disrupting the flow of power within a transmission system may have detrimental effects that can lead to system instability.

\subsection{Stakeholders' Investment Decisions}
{\ct

Existing, ongoing improvements to systems management for IP-based solutions help to protect critical cyber assets against electronic manipulations. Although promoting cyber-situational awareness helps grid operators to deal with extreme circumstances, preliminary pre-attack steps can add significant cyber-protection value as well. Cybersecurity of computerized automation systems for SCADA  networks governing the physical part of the grid can be a point of vulnerability if the critical cyber assets are not regularly audited for potential ``high-impact, low-probability'' attack events. Such audits should affect the prioritization of cyber asset investment. A tool for systemic security management with metrics quantifying detrimental effects of hypothesized attack scenarios has not yet become a standard procedure as part of the strategic investment planning for cyberinfrastructure. }
\begin{figure*}
\centering
\includegraphics[width = 18cm]{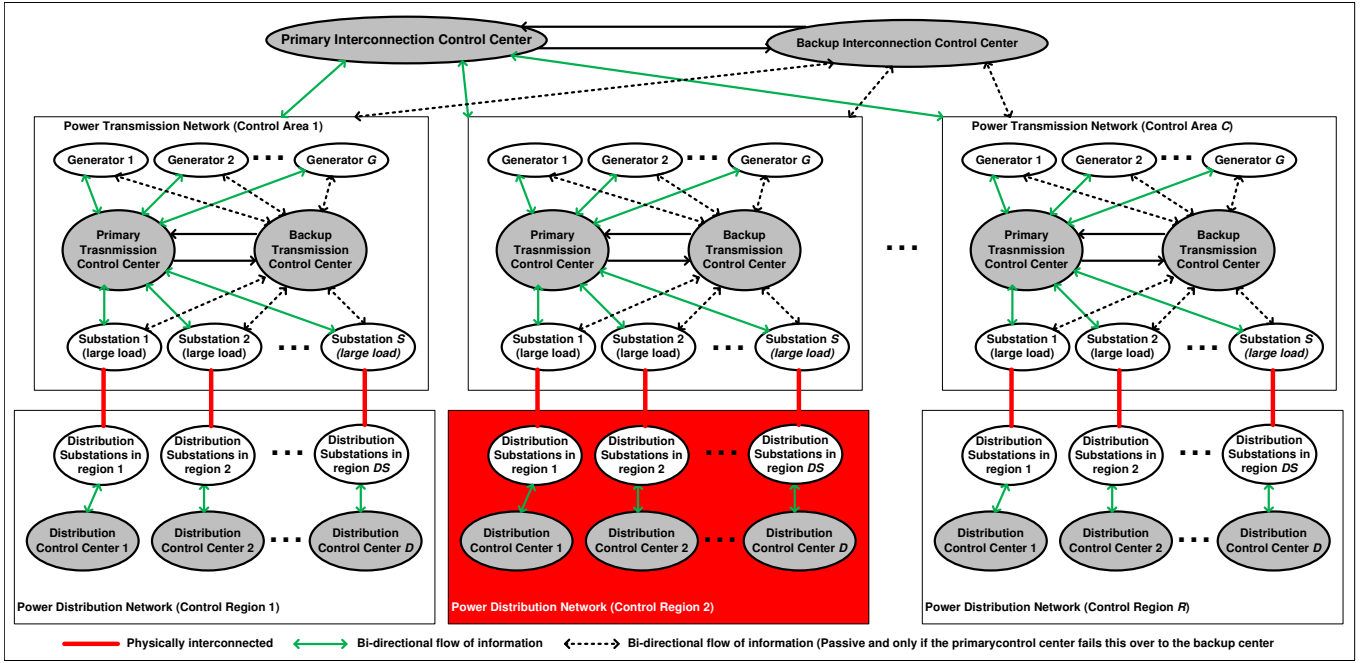}
  \caption{Generalized Wide-Area SCADA Network Connectivity Between Generation, Transmission, and Distribution Systems of a Power Interconnection.}\label{architecture} \vspace*{-5pt}
\end{figure*}

{\ct All substation-level and substation-connected networks are Wi-Fi or Worldwide Interoperability for Microwave Access (WiMAX)-disabled. In some extreme cases, the computers of the control system can be air-gapped by segregating from unsecured networks, with solely local cyber-physical system that has no communication at all to the rest of the world. This includes radio silence in which all employees are not allowed to bring their cellphone to the sites when they are in duty. The critical infrastructure network has the stringent rules against intrusion-based attacks in the sense of disallowing untrusted devices to the sites for the prevention of unnecessary pivoting to the control network. There may be other networks that are accessible by VPN or dial-in connections, which pose the real threats of external attackers. We also assume that the insiders would be a possible event that may assist the outsiders for plotting a cyberattack. Continued deployment of new communication infrastructure is changing risk management practices, both in terms of operation and planning.}

The United States Department of Energy (DOE) introduced an electricity subsector cybersecurity capability maturity model (ES-C2M2) tool as part of a Smart Grid Investment Grant (SGIG) project enhancement. This tool provides electric utilities and grid operators the ability to assess their cybersecurity capabilities, in order to prioritize investments in improving such capabilities. The progress of SGIG-implemented projects with cybersecurity plans (CSP) has been a catalyst for promoting awareness and specific control-framework risk-focused preparations with potential mitigation strategies. The United States National Institute of Standards and Technology (NIST) special publication 800-53 covers procedural steps in risk management that address security control selection for federal information systems. On-going research efforts are under way world-wide to improve cyber-security capabilities for critical infrastructures. The National Institute of Standards and Technology (NIST), International Electrotechnical Commission Technical Council (IEC-TC) 57, North American Electric Reliability Corporation (NERC), Critical infrastructure protection (CIP) drafting team, International Electrotechnical Commission (IEC) and Institute of Electrical and Electronics Engineers (IEEE) are working individually to develop different standards and regulation to enhance the cyber-physical security of power industries \cite{6,7}.

\begin{figure*}
\centering
\includegraphics[width = 18cm]{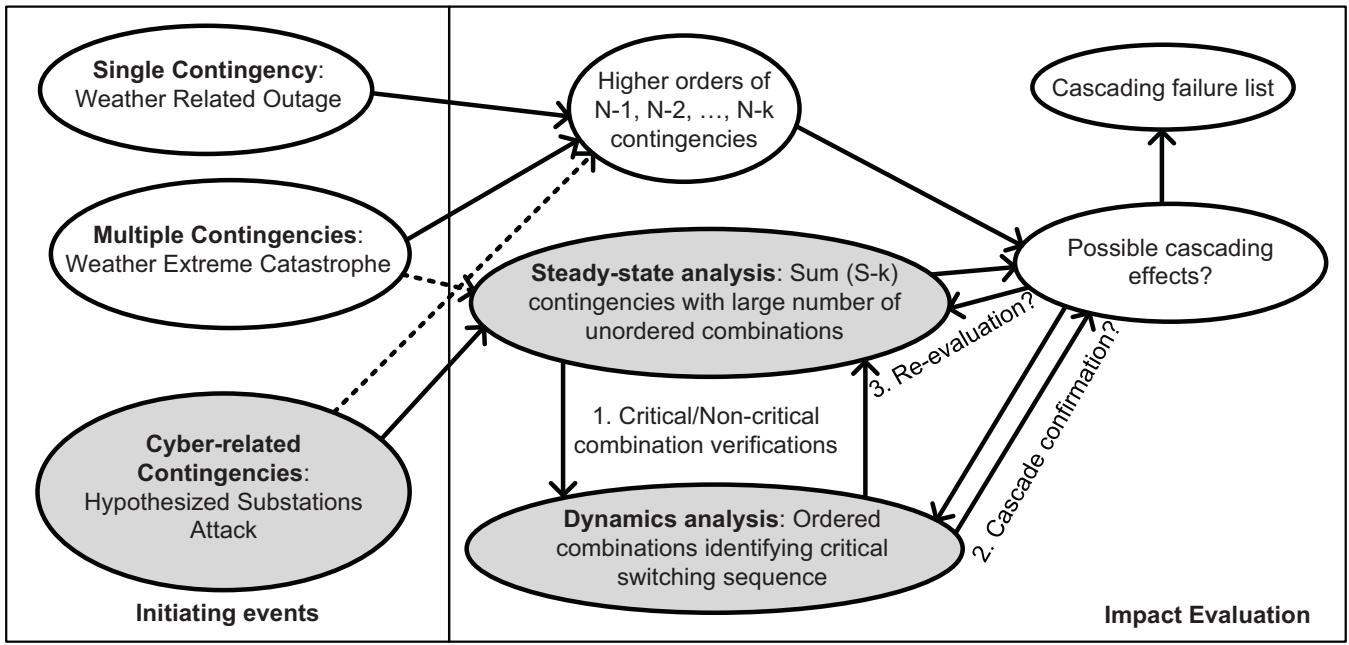}
  \caption{Conceptualization of Impact Evaluation, the `I' of RAIM \cite{9ad000004}.}\label{I-of-RAIM} \vspace*{-5pt}
\end{figure*}

The major contribution of this survey paper includes an extensive review of power control center risk assessment frameworks  {\ct which emphasize on the intrusion-based attacks to substation-level and substation-connected networks with a full spectrum of specific attack vectors that may disrupt operations, directly or indirectly. The physical impacts, which are the implications of cyber manipulations, can lead to widespread system instability that requires combinatorial evaluation and verifications of potential threats based on the plausible events from the access points as well as assistance by the insiders.} The contingency planning applications, i.e., ``what-if'' attack scenarios that remove components/substations from a power grid, ensures system stability in term of both stead-state and dynamical aspects in light of such attacks. This paper also describes a justification of cyber-based contingency evaluation with a thorough discussion of conceptual impact expansion of the {\cheewooi real-time monitoring, anomaly detection, impact analyses, and mitigation (RAIM)} framework  \cite{9ad000004} depicted in Fig. \ref{I-of-RAIM}. The remainder of the paper is organized as follows. Section II envisions the challenges and perspectives of the current state of the art. Section III describes intelligent attack scenarios on substations and their potential system impacts. Section IV introduces the I of RAIM which is a conceptualization of impact evaluation using both steady-state and dynamics analyses. Section V analyzes static and dynamical simulations cases using the IEEE 118-bus system for substation vulnerability assessment and system instability. Section VI provides concluding remarks.

\section{State of the art: Challenges and Prospectives}
{\ct The power grid has now emerged into one of the largest, most complex systems of human invention in all of history, involving tremendous communication bandwidth for the interactions between cyberinfrastructure and physical systems.} SCADA systems are an essential part of power communication infrastructures and play a central role in ensuring effective operations of bulk power systems. SCADA systems have helped to achieve new levels of system reliability and meet improved power quality requirements. The use of IP-based communication frameworks though, has brought about concerns over cybersecurity issues. As a result, the operational reliability of a power grid requires new methodological developments to align reliability goals with emerging risks of new communications technologies. Currently there are no comprehensive techniques and tools available to model and evaluate the hypothetical impacts of cyberattacks. The abrupt disruption or disconnections of nodes corresponding to load and generation can result in detrimental effects to the power grid. A US Government report published in 2007 reported several incidents of cybersecurity penetration in control system of different critical infrastructure \cite{1}.

\subsection{Power Control Center Framework}
Power infrastructure communication is integral to a nation's critical infrastructure \cite{RiskAssessment}. As early as the 1980s, the revolution of information communication technology (ICT) for power grid operation started changing how critical infrastructures are managed \cite{8}. {\ct As shown in Fig. \ref{architecture}}, ICT consists of generation local area network (LAN), transmission LAN and wide area network (WAN), distribution LAN and WAN, distributed generation LAN and WAN and Customer LAN networks. Different LANs are connected through public communication networks that are generally managed by telecommunication companies. There are often three hierarchical control centers: (1) a national control center, (2) regional control centers, and (3) local control centers. Each local control center collects real-time data from physical systems of substations and transmits that data to regional control center after processing the data. Distribution Control Centers manage local control centers for the largest substations in the distribution system. {\ct The Ukraine cyberattack compromised the distribution control center as highlighted with a red box in Fig. \ref{architecture}. This would impact significantly the overall grid operation in a global sense of potential cascading in case of generation-load mismatch}. Regional Control Centers are associated with managing high-voltage transmission lines and have supervisory control of all local control centers in that particular region. Regional control centers act as middleman between local control centers, distribution control centers and national control centers. Regional control centers mainly control transmission substations. National control centers play a vital role in power system operation and controls. Such centers control the extra high voltage (EHV) transmission system, coordinate the activities of regional control centers, and are responsible for overall power system reliability and stability. National control centers collect real-time data from regional control centers and perform the function of EMS, state estimator and central network management of the overall power system  \cite{10}. Studies by the North American Electric Reliability Corporation (NERC) have shown that a simulated cyberattack drill demonstrates an absolute possibility to bring down the US power grids \cite{SimulatedAttack}.

\subsection{Past, Current, and Future Applications of Contingencies}
The power grid is designed to withstand a single component outage (N-1 contingency), ensuring that operating limits are not violated by such outages \cite{NERCstd, NERCperform1}. Power system reliability evaluation includes the integration of individual substation operating states and contingencies which are measured in terms of power frequency and duration of substation equipment outage events. Failure criteria for substations and violation thresholds of system reliability are defined based on substation size, location and functionality within the system  \cite{34,35}. Literature review shows that there have been a number of blackouts caused by cascading failures of transmission lines and generating units in recent years throughout the world  \cite{32}. If a substation is de-energized, the change in power flow is compensated by other substations, which must have enough spare capacity to carry the excess power. If they do not, transmission lines and transformers of those substations will be overloaded and overcurrent protection will trip those components to avoid thermal damage. This event will initiate a cascading failure as the excess power is switched onto neighboring circuits, which may also be running at or near their maximum capacity \cite{33}. A probabilistic model can be used to estimate the cascading outages in high-voltage transmission network \cite{1664980} and online dynamic security assessment in an EMS environment \cite{648499}.

\subsubsection{Single Contingency}
Power system security is referred to the contingency analysis where an N-select-1 list of components is hypothesized as taken out of service to determine whether any such state results in a violation of voltage or power flow limits in a power grid \cite{PGOC}. In the 1970's, the traditional approach of steady-state contingency analysis is to test all contingencies, such as transmission line outage and loss of generation, that are predefined by system planner/operator’s experience and intuition \cite{4113452}. Inadequacies of this traditional approach were later addressed and new techniques proposed to perform exhaustive testing, including both primary and secondary contingencies. Contingency screening for fast real-time contingency analysis was proposed using modification of fast decoupled power flow algorithm \cite{141785}. An efficient contingency analysis method has been implemented to detect of flow violation for transmission lines  \cite{43179, MandatorySingleCont}.

\subsubsection{Multiple Contingencies}

To evaluate the contingency severity of removing any combination of substations from the system, an AC load flow method might be used \cite{4762167, NERCstdTPL003}. Multiple contingency is seen to have largely prepared reliable systems to survive disasters \cite{1425dfgf578,MultipleCont}. In North America, FERC has clarified that the list of the contingencies to be used in performing system operation and planning studies should include all the contingencies, N-1, N-1-1, N-2, as well as multiple contingencies \cite{MandatorySingleCont}. As required by NERC reliability standards, the power system after a contingency should return to a secure, reliable state within 30 minutes \cite{NERCstdTPL003}.

Multiple contingencies have been researched since the late 1970s \cite{4113452}. Since then, the main effort of such research is to reduce the computation burden caused by the tremendous number of contingency cases in bulk power system: the total number of N-k contingency cases is $N!/[k!(N-k)]$. Various screening and ranking techniques based on the theoretical approach and parallel computing techniques based on the simulation-based approach have been proposed and developed in the past five decades.

Conventional approaches related to the screening and ranking of contingencies are illustrated in \cite{4110642, 4110804, 5519401, MinorJournal1985, 192968}. After contingency studies for transmission planning were regulated by NERC in 2005 \cite{NERCstdTPL003}, research relevant to the NERC-compliance study have accelerated and the techniques for searching for critical/credible contingencies which consists of N-2 contingencies and N-1-1 contingencies have been developed in industry as well as academia \cite{NERCstdTPL003, 1425578}. Most of the approaches described by academia are based on network topology analysis \cite{1425578, 4162597} and nonlinear optimization heuristics \cite{1425578, 4162597} in terms of power planning perspectives. The hurricane Katrina disaster motivated power system engineers to consider the increasing risk of natural disasters and the necessity of online multiple contingency studies. Because multiple contingencies could lead to cascading failures, multiple contingencies have also been studied in terms of wide area monitoring and protection with sensors such as PMUs in the wake of wide-spread blackouts affecting North America and Europe in 2003, 2004, and 2006. Since then, multiple contingencies and consecutive large blackouts have been a frequently discussed topic in industry. Information about past blackouts have been shared by industries and academia all over the world every two years during CIGRE Paris session since 2006 \cite{CIGREsession2006, CIGREsession2008, CIGREsession2010, CIGREsession2012, CIGREsession2014}.

\subsubsection{Cyber-Related Contingencies}

As IP-based commu-nications infrastructure is the trend for future deployments, expecting only N-1 contingencies is no longer be meaningful for both security analysts and power engineers  \cite{NERCperform1}. As shown in Fig. \ref{I-of-RAIM}, a coordinated attack associated with compromised substations enables attackers to trip multiple generators, transmission lines, loads, or transformers nearly-simultaneously in a power grid, impairing system operating conditions. A more structured, integrated framework with high redundancy and defense mechanisms is required to face the challenges of intelligent coordinated cyberattacks, which can severely impact system operations \cite{6089026}. Violation of predefined thresholds of substation voltages, system frequency, and branch flows may lead to cascading failure and a system blackout \cite{6112807}.

As system loading levels vary over time, the criticality of each substation (node) can be different at different times \cite{1306711}. An approach to hypothesize multiple substation outages is proposed to presume that a set of combinations of IP-based substations are compromised by intruders and are electronically manipulated to abruptly isolate substations from the grid with disruptive switching actions \cite{Cybercontingency, cyberrisk}. Combinatorial substation outages are the cyber-contingency analysis that enumerates the worst-case scenarios. Since the solution space of the sum of S-select-k problem can be extremely large, a systematic elimination approach using power flow modules is used to validate each combination in order to capture the worst combinations \cite{CyberConJournal, Cybercontingency, cyberrisk, Impact}. This process eliminates insignificant combinations, enumerating from the first-level substation list of the RPM. While this approach may not be exhaustively enumerated; it can be further enhanced with prioritization of substation selection criteria. This contingency analysis is based on the relationship between substation critical cyber systems that have direct interaction with the physical power grid, i.e., the cyber assets that would have control capability to disconnect local components from the grid. Based on the conclusion of previous work \cite{Cybercontingency, cyberrisk}, manipulation of microprocessor-based relays on bus differential protection would have a detrimental effect, able to disconnect large numbers of components from the system. At minimum, hypothesized cyber attacks would occur at multiple substations, as attackers would be able to intrude to the S number of IP-based substations. Under this assumption, at least one or more substation outages would occur, depending on the number of substations that have been compromised \cite{Cybercontingency, cyberrisk, Impact}.

\begin{figure*}
\centering
\includegraphics[width = 16cm]{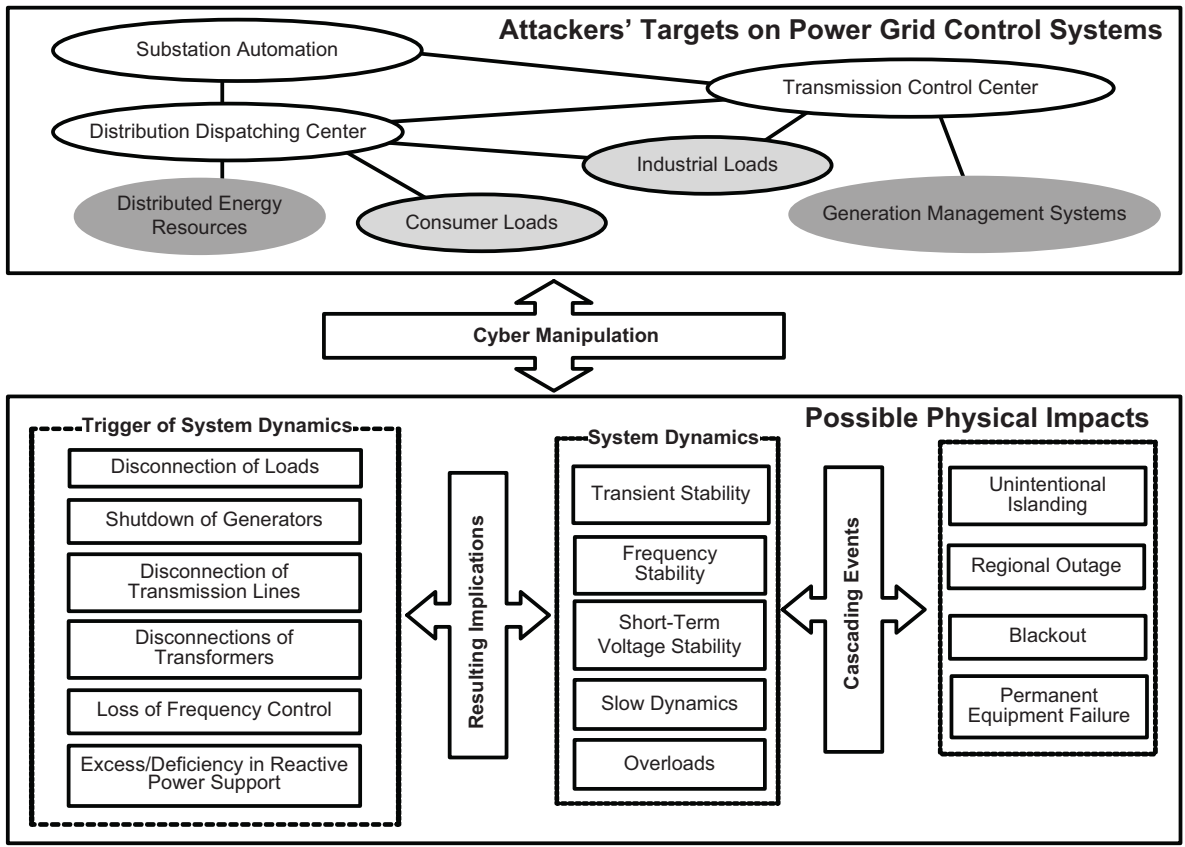}
  \caption{Cyberattack Implications.}\label{implicate} \vspace*{-5pt}
\end{figure*}
Preliminary investigation on system reliability and its resulting impacts has shown the effectiveness of the proposed algorithm with quantitative analysis on penetrated protective IED relays \cite{RealibityImpact}. Busbar differential relays might be of interest to attackers launching coordinated cyberattacks, since compromised busbars controls can disconnect more than one component from a system simultaneously. Such controls can be accessed by internal or external attackers in four different ways: physical attack, compromised communication channels, direct attacks on relays and changing the embedded program in relays \cite{6035612}. A small number of approaches exist to find critical substation combinations or collapse sequence of cascading failures, but none are in online environments, and all are applicable only to restricted numbers of combinations of substations \cite{4762167}. Since the event of simultaneous cyberattacks on 3 or more substations is rare, the worst case will be considered. Priority list 2 only evaluates the impact by de-energizing more than two substations that result in the impact factor of 1.0 and serves as a message to control centers  \cite{9000004}. This is a graph traversal algorithm that navigates the vertices of a graph first deep, then wide. This algorithm traverses from a root node in depth until a visited vertex is found \cite{algorithm}. It is commonly used in substations. Internal busbar faults are sensed by the busbar differential relay to trip all the breakers simultaneously connected to the busbar \cite{5325912}. {\cheewooi Cyber-based contingency analysis is a fundamentally new way to assess the system stability by considering all plausible attack vectors. These attack vectors can be any combination of these: (i) distributed denial of service (DDoS) \cite{6938963,6175560,6473865}, (ii) alter and hide (AaH) \cite{AaH}, (iii) data integrity \cite{6666849,6672740,6938963,6426269,6740883,6672731}, (iv) load altering \cite{6848210,ZhuHan}, or/and (v) disruptive switching \cite{6175560}.}

\subsection{Dynamics of Intelligent Cyberattack}
As shown in Fig. \ref{implicate}, {\cheewooi the existing major entities of bulk power systems have been upgraded with IP-based communication infrastructure over the past decades. The figure in gray highlights that are the generation and electrical loads. These interconnected entities have a profound impact by cyber manipulation, either locally, regionally, or globally. The} manipulation using a compromised local control system can impair system operation due to the potential impacts on the physical system. {\cheewooi The triggers of system dynamics, such as disconnections or abrupt shutdown of important elements within a power grid, can implicate the possibilities of system stability.}

{\cheewooi One of the worst-case scenarios is a widespread} cascading failure that will lead to a power blackout costing tens or hundreds of billions of the dollars to an economy as large as that of the USA. The importance of considering power system dynamics for cybersecurity issues has already been recognized. Power system dynamics can be significantly affected by network communication and control system infrastructure including generator controllers and protective relays. The establishment of a mathematical formulation for representing power system dynamics is a non-trivial task. Controllers and protections include non-linear behavior and discrete changes. In addition, any formulation must account for many interactions. There can be interaction between controllers and interaction between protection equipment. There can be interactions between controllers and protection, between controller and the grid, and between protection and the grid. This sub-section focuses on the review of the recent studies that are relevant to power system dynamics.

The recent research studies can be categorized as focused on either (1) Abnormal power system dynamic phenomena, or (2) Measurement of implementing cyber-physical security systems. Typically, an abnormal behavior of power system dynamics is classified into four phenomena that can result in a widespread power outage: voltage stability, frequency stability, transient stability and overload. The latest research studies cover the first three abnormal phenomena.

\subsubsection{Transient Stability}
Transient stability is examined using the undesired control of the semi-conductor-based reactive power compensators such as static var compensator (SVC) and static synchronous compensator (STATCOM). References  \cite{6666849,6672740,6938963} exhibit the possibility of being out-of-step due to biased or delayed operation of SVC or STATCOM. The fundamental idea is to represent the same dynamic behavior, even when the improper control parameters are tuned. Modification attack is assumed to be responsible for the undesired control. This vulnerability is relevant only when a system fault occurs near the reactive power compensator.

\subsubsection{Frequency Stability}
Frequency stability is examined using undesired control of Automatic Generation Control (AGC) or falsified load change data. Because falsified load changes have the same effect as an undesired control signal of AGC, the two attack scenarios can be treated as the same one. References \cite{6426269,6740883} exhibit the possibility of frequency collapse which results in significant frequency change, such as 3 Hz or more. The fundamental idea is to represent the same dynamic behavior when the wrong/improper control parameters of AGC are tuned. Data integrity attacks are assumed for the undesired control and the falsified load changes. The sudden loss of generation/loads can also cause frequency instability  \cite{6175560,6473865}. In this study, mono-directional frequency drop occurs because under-frequency relays are not considered, although the electric supply is less than the demand after the breaker trip. The fundamental idea is to create two or more isolated systems that cause a large mismatch of the power balance. A DDoS attack is assumed for the breaker trip.

\subsubsection{Short-Term Voltage Stability}
Short-term voltage stability is examined using the undesired control of stepwise change in active or reactive power outputs. Reference \cite{6672731}  exhibits the possibility of short-term voltage collapse which is caused by a significant voltage drop. The fundamental idea is to change active or reactive power output in order to generate a growing power swing oscillation and/or to have a shortage of reactive power support in the whole grid. In this study, transient stability problems seem to occur when a voltage collapse occurs. In the case of large networks, the short-term voltage collapse in entire power system could lead to an out-of-step condition in the entire network. Short-term voltage response is also examined using the non-operation of primary relay or unwanted operation of the back-up relay. Reference \cite{6672731}  exhibits the possibility of large voltage excursion. The fundamental idea is to enlarge the impact of the fault via non-operation of the primary protection or the unwanted operation of the back-up protection. However, the goal of this study does not represent blackouts, but to establish a complex cyber-physical system. Similar study includes using the undesired control of SVC caused by man-in-the-middle attack \cite{6965381}.

{\cheewooi
\subsubsection{Slow Dynamics}
Long-term dynamics is considered using the arbitrary load change by jamming the pricing signal in the electricity market. References \cite{6848210,ZhuHan} studies the possibilities of unwanted slow dynamics caused by the delayed and distorted data-centric attack, which eventually causes the degradation of the controller performance and the negative impact on any kind of the power system stability. Smart meters in electrical distribution network which utilize wireless communication such as WiMAX is assumed to be used for this scenario and the jamming attack is applied to the electricity market in order to jam the power price signaling over a large area such as the load center. Such manipulation of the electricity market via the data-centric attack (or the false data injection attack) can bring the attacker to the profit and cause the significant impact on the stability of the power system.
}

\subsection{Implementation of Cyber-Physical Security System}
{\ct In order to implement a cyber-physical system including power system dynamics, two streams can be classified: (1) \emph{Simulation-based} approach, and (2) \emph{simulator-based} approach. The first approach is often used for large network studies, while the second approach is often used for small network studies because of the space limitation of the test bed or the test facilities. Therefore, such simulators are suitable for representing frequency stability studies because the large network representation is not necessary. The transient stability study is not examined using any simulators or any test beds. In order to evaluate the performance of the test bed, small signal stability or steady-state stability, instead of abnormal power dynamics phenomena, is selected in \cite{6848210}. The second approach can represent a relatively large network such as IEEE 39 bus and IEEE 118 bus systems. On the other hand, a mathematical approach of establishing a hybrid system is proposed and its performance validated using a simple network without assuming the specific cyberattack \cite{6062375,6344956}.}

\section{Intelligent Attack Scenarios on Substations and System Impacts}
{\ct A distributed denial of service attack (DDoS) harnesses the power of agents to disrupt communication infrastructure that can interrupt the availability of services in substation networks. Possible consequences include preventing information exchange among relays and programmable controllers. As a result, information processors and IEDs are not aware of the current status and fault condition of the system. As this blocks alarm and trip signals, a DDoS attack on transformer and line protection schemes of an IED does disable the functions of differential protection and directional over-current protection. This is similar to a situation where a malfunctioning relays might not perform the operation as they should, based on the original settings of substation logic controllers.}

Upon a successful intrusion to a substation network, attackers can plot for traffic manipulation using their domain-specific knowledge. The required cyber-physical security understand-ing between the standardized communication protocols and the connection with the physical devices is crucial to maximize the impact of attack. The attacker would have to understand software setup, and understand how device addresses map to a user interface in power control centers. The obvious manipulation is to add delay for each signal that affects not only the protection scheme but also SCADA functionalities at the control center. Blocking a trip signal for certain time will delay the breaker-trip in a faulted section of transmission line and can have an adverse effect on system operation. This section enumerates credible intelligent attack strategies upon their success intruding to a substation network.

\subsection{Scenario 1: Transformer Overloading}
Transformers are the critical components of power sub-stations, and play a crucial role in transferring power from electrical generators to loads through other geographically dispersed substations. Failed high-voltage transformers can require very long times to replace. A number of protection schemes are used to protect a transformer from various fault conditions. Transformer life is limited because of aging of insulation. According to IEEE Std C57.91-1981, the expected life of power transformer is 180,000 hours or approximately 21 years. IEDs are programmed to calculate the time left before a possible trip due to loss of insulation life \cite{23}. Several hypothesized scenarios present how a potential cyberattack can cause damage to substation equipment. Studies show that there are several ways to get into a substation communication network. In this section we consider the hypothesis that an attacker has the total control of substation communication network and have access to IEDs which are the source of substation and IED configuration descriptions. After having access to SCD and CID files, an intruder can plan for a devastating attack on substation operations. SCD and CID files are the source of substation topology, protection schemes, back up protection schemes and crucial settings for different IEDs.

\begin{figure*}
\centering
\includegraphics[width = 15cm]{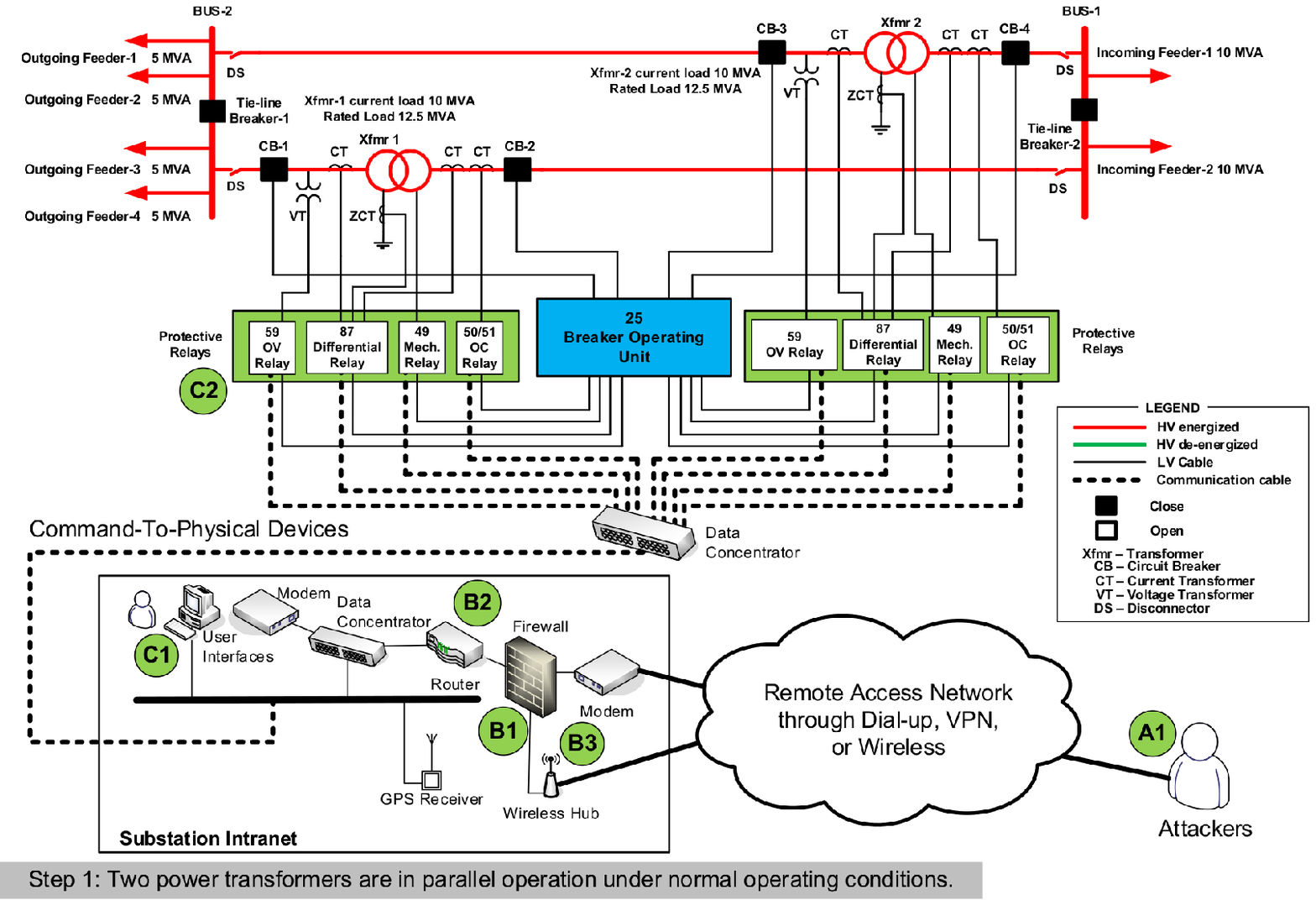}
  \caption{Case 1: Cyber-Physical Security of a Power Substation.}\label{drawing3} \vspace*{-5pt}
\end{figure*}

\begin{figure*}
\centering
\includegraphics[width = 15cm]{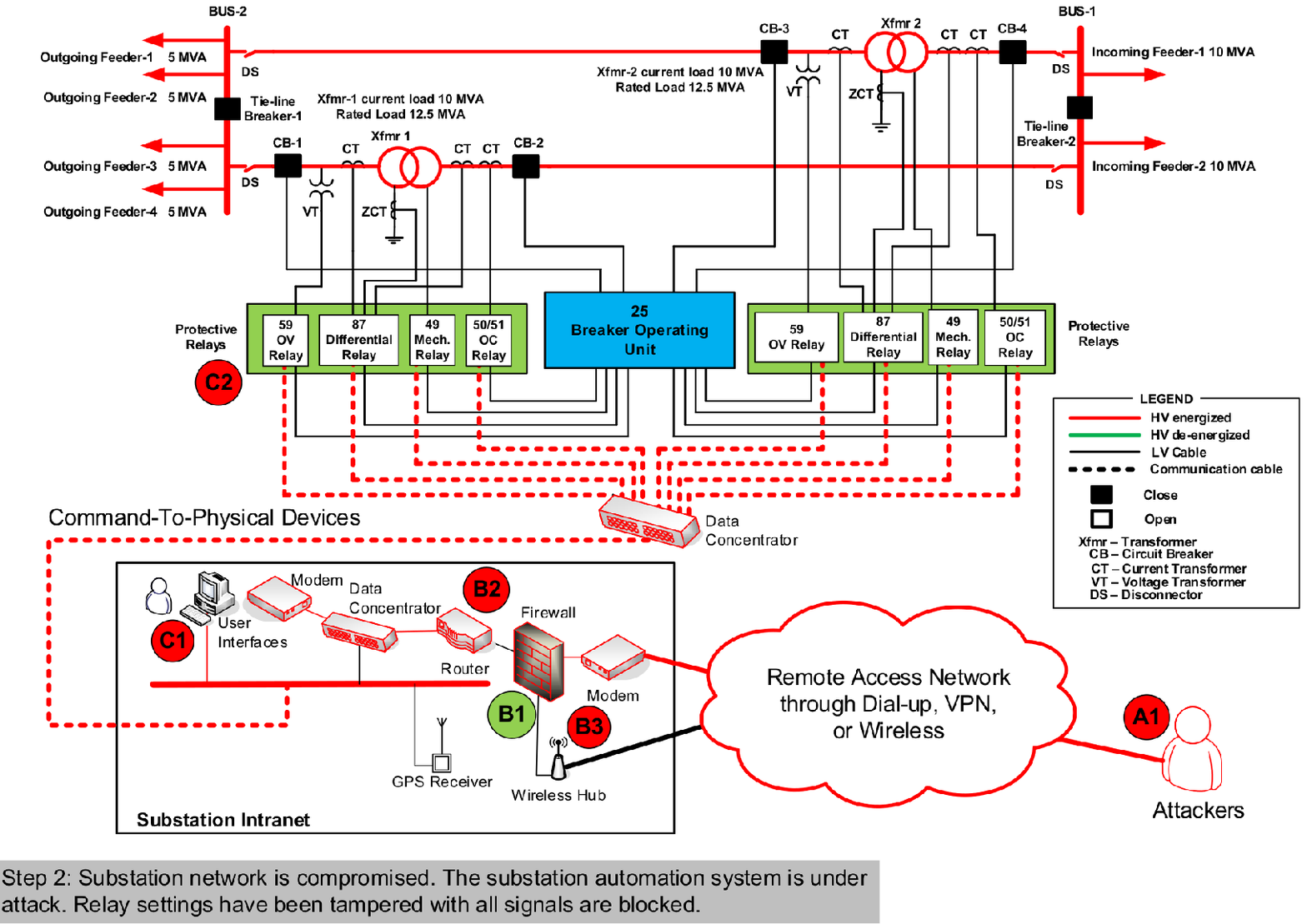}
  \caption{Case 2: An Attacker Gained Administrative Privilege to Local SCADA Network.}\label{drawing4} \vspace*{-5pt}
\end{figure*}

Fig. \ref{drawing3} shows a typical substation configuration where Transformer-1 (Xfmr-1) and Transformer-2 (Xfmr-2) are in parallel operation and sharing the total load of 20 MVA. This is a common practice of using at minimum two transformers to share a load. Each Transformer rated power is 12.5 MVA. In the diagram, the illustrated transformer current load factor is 80 percent. Two incoming feeders and four outgoing feeders are connected in BUS-1 and BUS-2 respectively. Fig. \ref{drawing3} shows that Xfmr-1 and Xfmr-2 are protected by differential protection, thermal protection and over-current protection. A simple substation communication network is depicted, using an information processor, a gateway, a modem, server, a remote I/O unit and a phasor data concentrator. Fig. \ref{drawing4} depicts the ethernet gateway, modem, and wireless link which are the access points used by a cyber attacker to breach the substation communication network. Once the attacker gains access to the communication network, (s)he can access substation SCD and CID files. To execute transformer overloading protection failure, the attacker need to modify the overcurrent and thermal protection schemes. It is assumed that intruder will reset the threshold values of different relay trip settings to much higher values so that the relay will not act to trip a breaker during a typical fault condition. For example, if the normal breaker trip setting for overcurrent protection is 150\% overloading and an attacker increases this value to 300\% then the breaker will not trip for less than 300\% overloading. Similarly, thermal protection will not work if the breaker trip setting is changed to a much higher value. DDoS attack or traffic manipulation can block all kinds of alarm and trip signals from bay level to station level. As a result, local and regional control centers will not be able to receive any real-time information, such as transformer trips, alarms, equipment status, other system status, from process and bay levels.

\begin{figure*}
\centering
\includegraphics[width = 15cm]{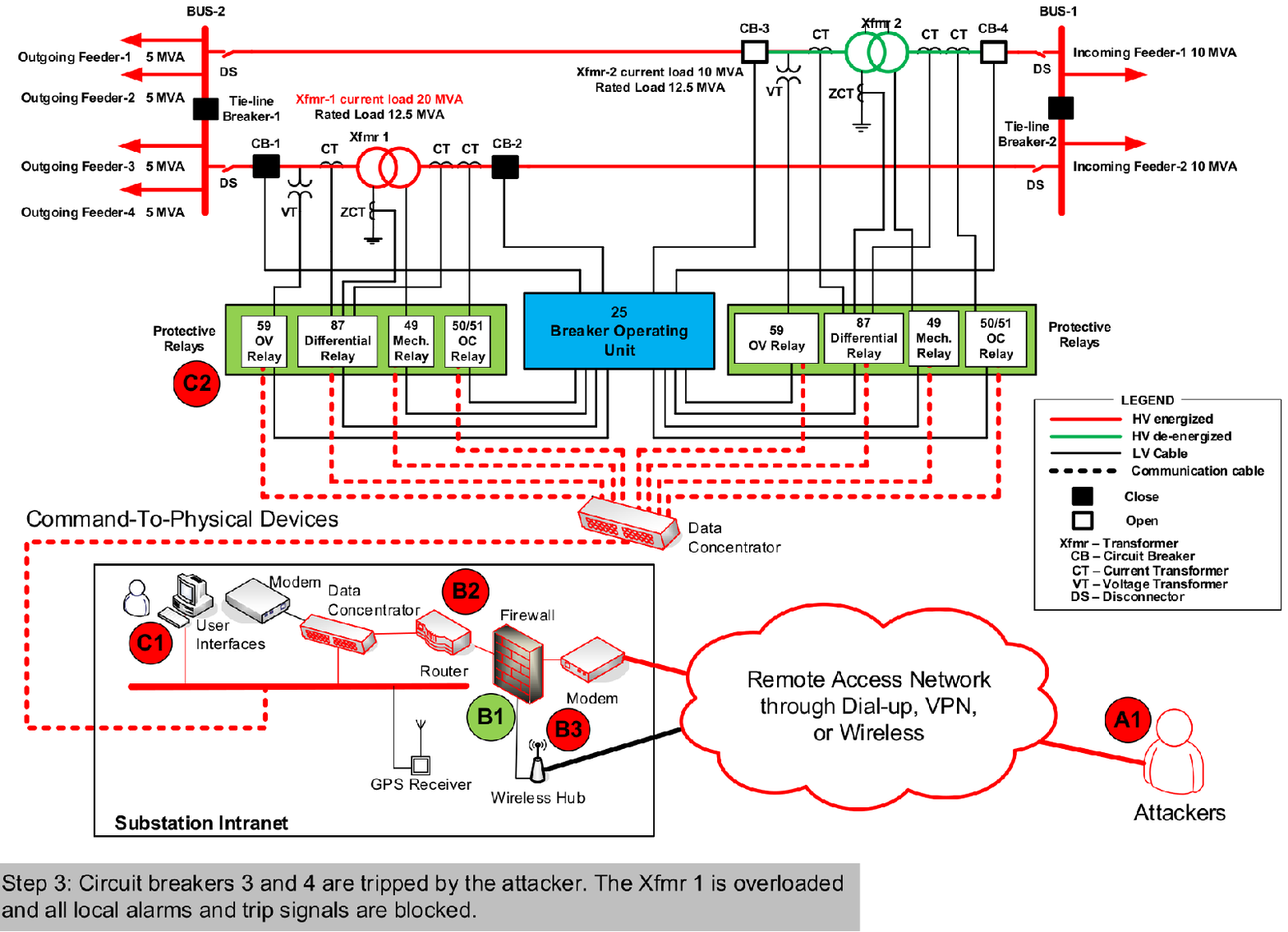}
  \caption{Case 3: Attack Execution to Overload a Power Transformer.}\label{attackexe} \vspace*{-5pt}
\end{figure*}
Once trip settings of overcurrent and thermal protection relays have been modified and the alarm signals from relays and other IEDs have been blocked, the attacker is ready to overload one of the transformers. As shown in Fig. \ref{attackexe}, the attacker can open circuit breaker-3 (CB-3) or circuit breaker-4 (CB-4) or both to shutdown Xfmr-2. Since all other breakers except circuit breaker-3 and circuit breaker-4 are in a closed position, there is no power interruption in that substation. As a result, the Xfmr-1 will carry the total load of 20 MVA and hence it will be overloaded by 160\%. In this overload condition, the transformer’s temperature will increase rapidly. This will adversely impact transformer insulation and transformer life \cite{24}.

Studies on transformer show that most of the transformer failure is directly related to insulation deterioration with time. IEEE standard states that reduction of transformer insulation is a function of time and temperature \cite{26}. Power loss $(I^{2}R)$ of the winding and the stray loss in the tank, core and other metallic structures are the main source of transformer heat generation. Mechanical strength of transformer insulation deteriorates with time and becomes brittle when it is exposed to higher temperature \cite{24}. According to IEEE Standard C57.91-1995 the overall life of a transformer is equal to life of insulation. the normal life expectancy of a power transformer is 20.55 years when it is assumed that transformer
operates  at constant $30\,^{\circ}{\rm C}$ average ambient temperature and the hottest-spot winding temperature will not exceed $110\,^{\circ}{\rm C}$ \cite{25,26}. Since the temperature
rise of transformer is not uniform, the hottest-spot temperature is considered to find out the insulation aging factor $(F_{AA})$ \cite{24}. IEC standard states that relative aging rate is doubled for every $6\,^{\circ}{\rm C}$ increment of windings hottest-spot \cite{25,26}. IEEE Guide for Loading Mineral-Oil-Immersed Transformers summarizes the following risks of
transformer overloading and its implications of potential equipment damage \cite{26}.
\begin{enumerate}
\item The evolution of free gas from insulation of winding and other metallic, structural parts of transformers reduce dielectric strength.
\item Operating at high temperature will reduce the mechanical strength of both conductors and structural insulation. This reduction of strength can cause permanent damage to transformers if the mechanical forces during transient overcurrent faults exceed the reduced mechanical strength.
\item Thermal expansion of conductors, insulation and other metallic parts can permanently deform transformer bodies.
\item Rapid thermal expansion of transformer oil will increase the pressure inside the transformer and result in oil leaking and insulation failure.
\item Current transformers installed in bushing and reactors are also at risk due to higher temperature.
\item Increased resistance of tap changer contacts can create a localized high temperature region which could result arcing and gas evolution.
\end{enumerate}

{\ct Two types of emergency overloading are described in IEEE standards:
\begin{enumerate}
\item \emph{Long-Term Emergency Overloading:}
This kind of emergency condition arises when system components are out of service for an extended period. IEEE Standard transformers tolerate long-term, steady state overloads with some loss of transformer expectancy. In \cite{26}, acceptable long term overloading is defined as a temperature rise of hottest-spot windings limited to $120\,^{\circ}{\rm C}$ to $140\,^{\circ}{\rm C}$, where the top oil temperature does not exceed $110\,^{\circ}{\rm C}$ any time. This type of overloading may occur two or three times over the normal life of a standard transformer and each incident can last several months \cite{26}.

\item \emph{Short-Term Emergency Overloading:}
This emergency condition sees a standard transformer carry a maximum 200\% overload for a short period, limited to a maximum of 30 minutes. In such conditions, the standard allows a transformer’s hottest-spot winding temperature to reach no more than $180\,^{\circ}{\rm C}$, and requires that the peak oil temperature not exceed $110\,^{\circ}{\rm C}$ any time. The risk of short term overloading is more severe than long term overloading. Only one or two occurrence of such overloading are recommended over a normal life of a standard transformer  \cite{26}.
\end{enumerate}

\begin{figure}
\centering
\includegraphics[width = 9cm]{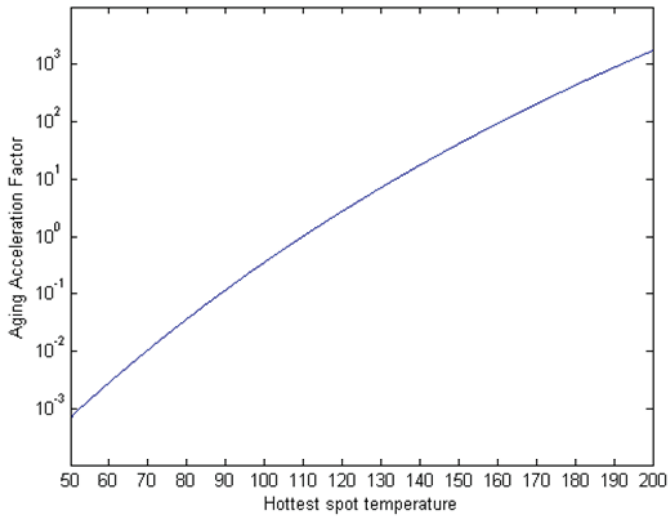}
  \caption{Aging Acceleration Factor Based on $110\,^{\circ}{\rm C}$.}\label{FAA} \vspace*{-5pt}
\end{figure}
In the hypothetical attack scenario, the load of Xfmr-1 is doubled instantly and the transformer’s temperature will increase rapidly. If Xfmr-1 is an IEEE-compliant transformer, a 160\% overloaded condition will cause the transformer’s hottest spot temperature to reach up to $160\,^{\circ}{\rm C}$ \cite{25} where transformer insulation aging acceleration rate $(F_{AA})$ is expressed by $F_{AA} = \exp{\Big(\frac{15000}{383}-\frac{15000}{\theta+273}\Big)}$.

Fig.  8 shows that $F_{AA}$  changes exponentially with respect to temperature. It can be noted from Fig. \ref{FAA} that the value of insulation aging acceleration is 1 and 92.1 at $110\,^{\circ}{\rm C}$ and $160\,^{\circ}{\rm C}$ respectively. It shows that $F_{AA}$ at $160\,^{\circ}{\rm C}$ is 92.1 times faster than at $110\,^{\circ}{\rm C}$. IEEE Standard C57.91-1995 gives the percent loss of transformer insulation life due to continuous operation above rated hottest-spot temperature \cite{26}. It is noted in IEEE standard that a transformer will lose 1\% and 4\% of its normal life due to continuous overloading for 1.96 hours and 7.8 hours respectively at $160\,^{\circ}{\rm C}$ hottest-spot temperature. If this situation continues transformer insulation might fail in less than 48 hours. Under extreme circumstance, a transformer can plausibly be overloaded by more than 200\% when three transformers are in parallel operation in this kind of attack. The hottest-spot temperature may rise more than $200\,^{\circ}{\rm C}$ in this type of attack on a 3-way parallel transformer \cite{25}. According to IEEE standards $F_{AA}$  is approximately 1723 at $200\,^{\circ}{\rm C}$ hottest-spot temperatures. This situation results 4\% loss of transformer life for every 25 minutes continuous operation. Such a transformer might be destroyed in less than 10 hours if this overloading condition continues.

\begin{figure*}
\centering
\includegraphics[width = 18cm]{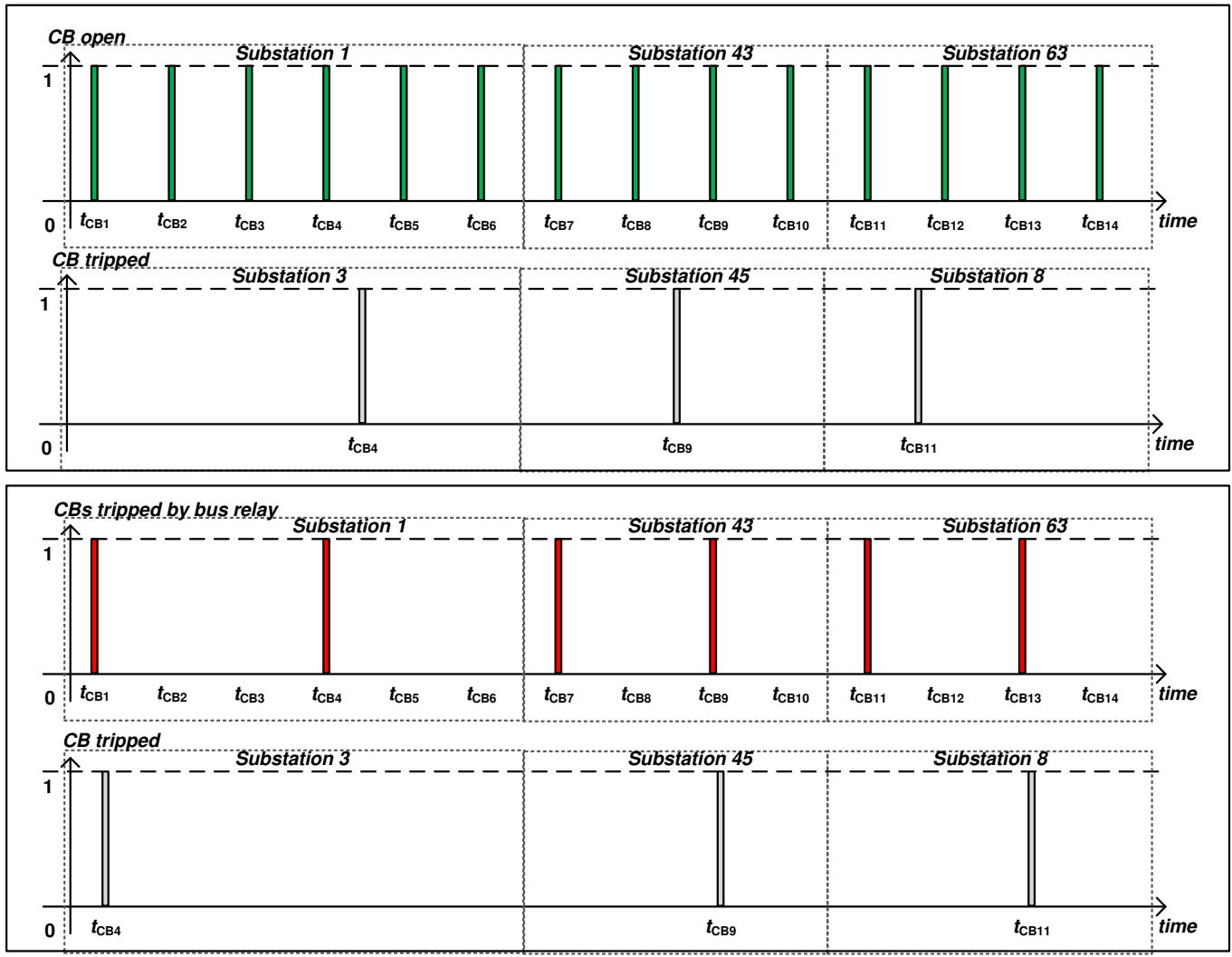}
  \caption{Switching Attacks Upon Compromised Substation Networks.}\label{switchattack} \vspace*{-5pt}
\end{figure*}

\begin{figure*}
\centering
\includegraphics[width = 15cm]{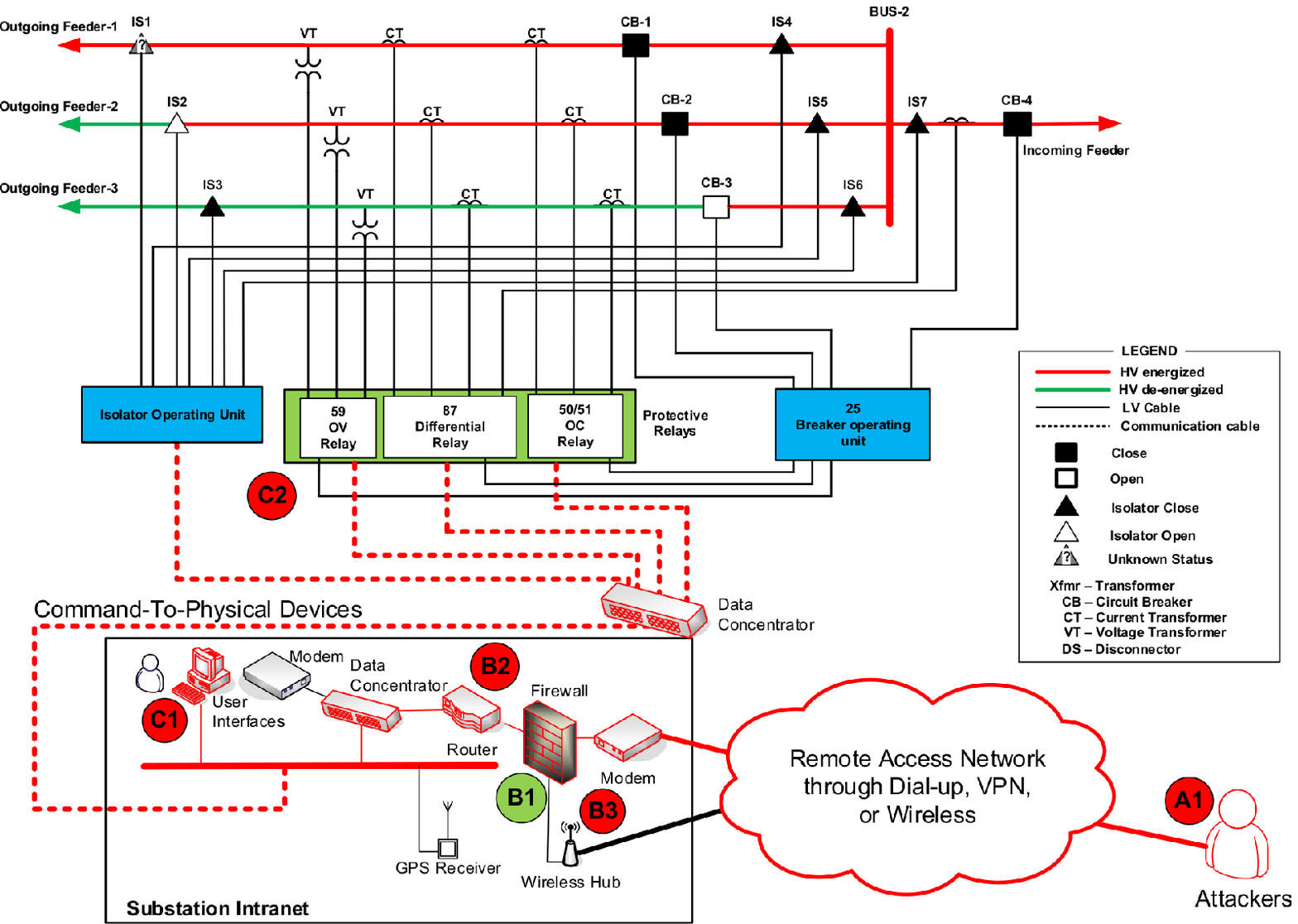}
  \caption{Disruptive Switching Action of Circuit Breakers and Isolators.}\label{drawing6}
\end{figure*}
}
In this hypothesized attack scenario, some other issues need to be considered for insulation aging:
\begin{enumerate}
\item During overloading, gas may form inside transformer tank and affect the integrity of insulation.
\item The hottest spot temperature may vary from manufacturer to manufacturer and the rating of transformer.
\item Tap changer operation during overloading may cause additional damages.
\item Cooling fan failure of the transformer radiator expedites the temperature rise.
\item It is very unlikely that the overloaded transformer in such an attack is brand new. Thus, it may take less time to destroy the insulation rather is calculated above.
\item An ambient temperature of more than $30\,^{\circ}{\rm C}$ will increase the rate of insulation aging.
\end{enumerate}

\subsection{Scenario 2: Disrupting Switching Executions for Circuit Breakers and Isolators }
In this hypothesized scenario, an attacker focusses on damage to circuit breakers or isolators to interrupt the normal operation of a substation. As depicted in Fig. 9, switching sequential attacks can result in power substation outages. A switching attack upon compromised multiple substations can initiate system instability that can trigger widespread outage in a power interconnection. The assumptions for this scenario are as follows:
\begin{enumerate}
\item Unmanned, remote substations have no video surveillance and minimal physical security. Isolators are not interlocked with circuit breakers, and can be operated from compromised computers.
\item An attacker is capable of operating circuit breakers and isolators using configuration information available in the relays and user interface of local control panel.
\item All digital signals reporting circuit breaker and isolator operation are blocked. As a result, the regional control center will not receive any authentic real-time information regarding their use.
\end{enumerate}

Fig. \ref{drawing6} shows that one incoming feeder and three outgoing feeders are connected to a busbar, and a total seven isolators and four circuit breakers are connected to protect the feeders. The hypothesis is that a cyber intruder has access the communication network of the substation to commence continuous switching actions to take one or more pieces of equipment out of service. Although circuit breakers and isolators are disconnecting switches, a circuit breaker is designed to operate while lines are energized, whereas an isolator is designed for use while de-energized. The following are the main causes of circuit breaker failure due to disruptive switching:
\begin{enumerate}
\item Mechanical failure of spring charged motor
\item Insulation failure
\item Loss of arc interrupting medium, e.g., SF6 gas, air pressure, or oil pressure
\end{enumerate}
Dielectric break down of the air gap between contacts causes arcing during opening and closing of circuit breakers and isolators. Arcing and thermal ionization results in high temperature on contact surfaces, causing contact welded and erosion. Arcing may persist for 4–6.5 seconds, which might increase contact resistance by several multiples [81]. Isolators are more vulnerable to arching than circuit breaker, because isolators have no arc-extinguishing medium. The erosion of contact materials depends on opening and closing times, which depend in turn on motor charged spring operation \cite{28}. Increased resistance of eroded contacts causes heat during conduction. It is recommended that the temperature of contact surface not exceed $110\,^{\circ}{\rm C}$  for copper and aluminum conductors \cite{30}.

\begin{figure*}
\centering
\includegraphics[width = 16cm]{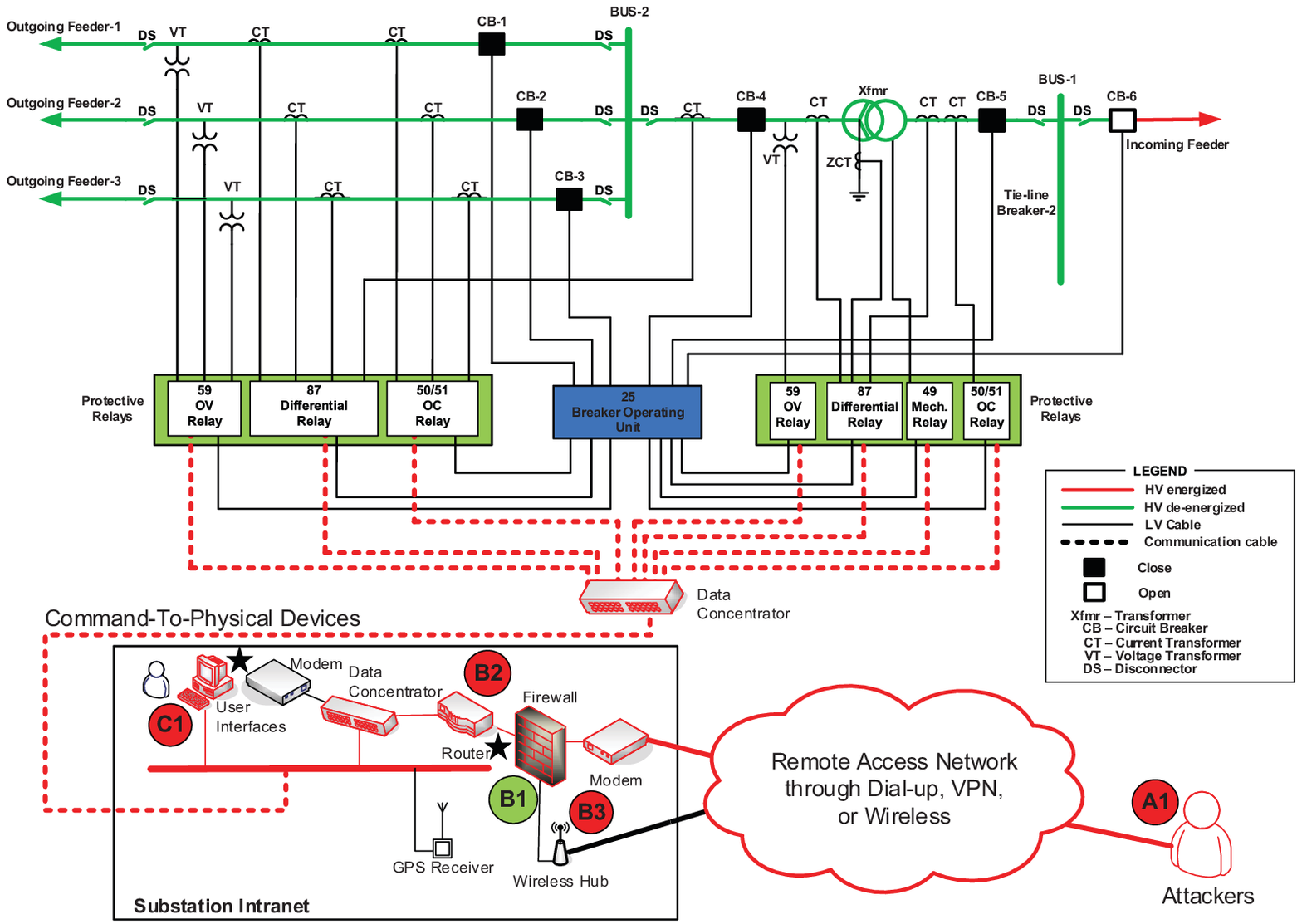}
  \caption{Entire Substation Outage by Opening the Circuit Breaker 6.}\label{drawing7} \vspace*{-5pt}
\end{figure*}

Study shows that mechanical inadequacy of contact breaking is the cause of 75\% of circuit breaker and isolator failures. The failure of this critical equipment increases with the voltage level and number of operations \cite{27}. The failure of isolators is associated with the following reasons:

\begin{enumerate}
\item Arcing contact failure
\item Main contact failure
\item Alignment failure
\item Motor failure
\end{enumerate}
The most common mechanical endurance failure is spring charging motor failure. A motor limit switch is responsible for motor relay energization to store potential energy in springs which are prone to fail with excessive circuit breaker operation. The failure of charging springs is expressed as circuit breaker failure. In case of isolators, eroded contact surfaces results in unacceptable heat during conduction, which deforms the contact surfaces.

\subsection{Scenario 3: Entire Substation Outage}

This hypothesized scenario represents entire an substation out-age and associated contingencies. A substation can be treated as a super node of a power system electrical network where multiple incoming and outgoing transmission lines/feeders are connected. De-energization of a critical sub-station may lead to a catastrophic cascading failure of entire system. In this study, the term ``substation outage'' refers to the complete electrical disconnection of a substation from the power grid, including de-energization of busbars, transformers and transmission lines. A substation outage causes topological changes to a power grid network as a number of transmission lines/feeders and busbars are isolated from the system.

Fig. \ref{drawing7} shows a typical distribution substation topology where an incoming feeder and three outgoing feeders are serving the loads. Outgoing feeder-1, 2, 3 and the incoming feeder are protected from any internal and external faults by circuit breaker-1, 2, 3 and 6 respectively. A successful intrusion into relays and local user interfaces enables an attacker to operate any of the circuit breakers in the substation. Substation outage can be caused in a number of ways, depending on busbar and breaker arrangements. The outage attack illustrated in Fig. \ref{drawing7} opens either all circuit breakers simultaneously, or circuit breaker-6 only. Since there is a single incoming feeder, opening breaker-6 ensures an entire substation outage while circuit breakers 1, 2, 3, 4 and 5 remain closed. In both case all busbars and transformers will be de-energized. The green lines in Fig. \ref{drawing7} represent the de-energized region of the substation. Knowing substation topological configuration prior to executing a cyberattack will assist an intruder in isolating the substation from the power system. Such an attack on a transmission substation is more critical than on a distribution substation since the former carries a larger amount of power and is generally responsible to maintain connectivity with multiple other substations in the grid.

Power system reliability evaluation includes the integration of individual substation operating states and contingencies, which are measured in terms of power frequency and the duration of substation equipment outage events. Failure criteria for substations and violation thresholds of system reliability are defined based on substation size, location and functionality within the system \cite{34,35}. A literature review shows that there have been a number of blackouts caused by cascading failures of transmission lines and generating units in recent years throughout the world \cite{32}. If a substation is de-energized, power flows through other substations, which must have enough spare capacity to carry the excess power. If they do not, transmission lines and transformers of those substations will be overloaded and overcurrent protection will trip those components to avoid thermal damage. This event will initiate a cascading failure as the excess power is switched onto neighboring circuits, which may also be running at or near their maximum capacity \cite{33}.

When a substation fails due to cyberattack, power flows shift to adjacent lines. This kind of event may overload the adjacent transmission lines and eventually the overloaded lines will be tripped to avoid transmission line damage. Redistribution of power flows through a transmission network may overload multiple lines. The number of overloaded transmission lines will be increased if this situation continues, eventually resulting in a cascading outage \cite{33}.


\section{Impact Evaluation}

A hypothesized impact study evaluates the plausible consequence of cyberattacks and anticipates system behaviors based on a certain operating condition. Such prediction of potential cascading outages and failures can be utilized to derive metrics that quantify the attributes of a case study. This requires a validation of impact credibility with respect to methods used in steady-state and dynamic simulations. As shown in Fig. \ref{I-of-RAIM}, a conceptualization of impact evaluation is proposed to handle the combinatorial nature of a cyber-related issues associated with (1) critical/non-critical combination verifications, (2) cascade confirmation, and (3) re-evaluation.

\subsection{Critical/Non-Critical Combination Verifications}
As depicted in Fig. \ref{I-of-RAIM},  this verification involves steady-state and dynamic analysis of the hypothesized components and substation outages. This check point is used to determine inconsistent simulation outcomes from both dynamic and static modules, and to reconcile the difference through cascade confirmation and probability re-evaluation through an adjustment of parameters. For example, a hypothesized substation outage would result in power flow diverged that may not necessarily reflect the similar outcome in the dynamic simulation. This could happen where a power flow simulator shows no symptom of cascading failure but dynamic study indicates otherwise.
\subsection{Cascade Confirmation}
The cascade confirmation pro-posed here is to determine the coherency of relative angle and frequency under certain switching permutation. The studies will include determination of the number of permutations that is deemed conclusive and adequate corresponding to sequential contingencies. The effective pre-screening of sequential contin-gencies would include substation dependencies and the practicality of concurrent cyberattack to abruptly disconnect components/substations out of the grid. The challenges here adequately represents an unordered combination of outage from the steady-state power flow module.

\subsection{Re-Evaluation}
Re-evaluation often occurs when both simulators demonstrate significant deviation of outcome that requires an adjustment. Such errors can be related to either of the simulators where specific handling is necessary. The discovery of tuning between the two modules can significantly reduce of the discrepancies and will strengthen the verification and credibility of the hypothesized scenarios. The results depend on the size of a power system. Fig. \ref{criticalnon} enumerates all critical and non-critical cases and illustrates cascaded evaluation with steady-state rapid screening to dynamic simulation to verify the catastrophic scenarios. {\ct The divergence evaluation in a power flow model determines the criticality of hypothesized substation outages. However, the results of steady-state evaluation may not consistently demonstrate the same outcome in the sense of potential grid instability of cascading implication.

{\cheewooi
Fig. \ref{criticalnon} describes the dependencies of steady-state simulation screening with a probability of the critical contingency $P$ that is derived as 0.1\%, while the probability of the non-critical contingency is derived as 99.9\%. The summation of probabilities at each level (steady state and dynamics) is always equal to 1.0. These probabilities are the statistical numbers that will be established to determine the total number of convergent and divergent cases. The steady-state approach would largely eliminate the criticality of problematic scenarios which may not be always true. The latter part of dynamic simulation is to examine the consistencies of both methods whether they are stable or unstable from the power system stability point of view. This may not always be the same between the both. The verification of dynamic simulation could reveal the criticality of non-critical contingency which is statistically derived. As shown in the figure, the critical combinations with probability 0.1\% based on steady-state simulation will be examined with dynamic simulation that may result in 73.5\%. On the other hand, the dynamic simulation verification can also reveal that the probability of the critical contingency that is statistically derived by 4.7\% from the steady-state probability of the non-critical contingency, which is 99.9\% in most cases. Computationally, the steady-state approach is much less expensive than the dynamic simulation verification while the latter approach has a higher degree of accuracy with detailed description of system behaviors.
}

\begin{figure}
\centering
\includegraphics[width = 9cm]{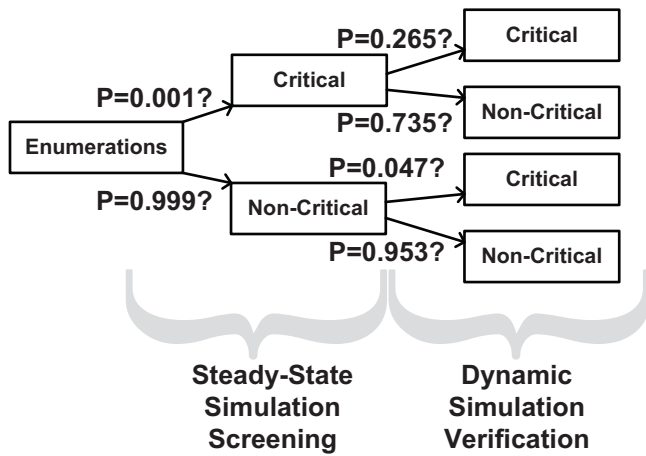}
  \caption{Enumerative Grouping for Critical and Non-Critical Cases.}\label{criticalnon} \vspace*{-5pt}
\end{figure}

In general, steady-state analysis may not always reflect a potential issue of potential cascading, especially the transient states are not sequentially captured one after another subsequent outages. The stability evaluation of cascading outages is a dynamic-security analysis with variables in transient status such as time period, frequency and voltages instability, generation-load imbalance, and metrics of cost-benefit justification of catastrophic impacts \cite{cascading1,cascading2,cascading3,cascading4}. For example, the mismatch of generation and load in power flow evaluation model can be numerically balanced on the ``reference'' node, where physically, there is no slack bus in the practical world. The imbalance of generation and load can only be achieved by adjusting generation output and/or shedding load. The time transition of one state to another is not sequentially captured in the steady-state power flow model -- the power flow evaluation solely addresses the divergence-convergence issue which is directly related to the initiation of voltage, generation, bus type, and topology of power system.  The time interval is critical in the dynamic analysis with cascading outages. Sequential cascading events can be induced in a bulk power system after an initial cyberattack that can range in an interval of $5$s to $15$s for a switching action initiated by attacker automation tools. The operation condition of power system should reflect the transition states that may lead to instable or not in term of generator ramping up/down or load shedding as well as the influence of various protection scheme deployed on any microprocessor-based components within a substation-level network. Consequently, the active power flow of each transmission line needs to be updated during each time interval. Moreover, under certain circumstance, the models of steady-state analysis may be inadequate to detail the sequential disruptive switching permutation under a substation specific scheme, e.g., one-an-half bus-breaker model because the IEEE test case may be simplified in most of the steady-state analysis appeared in the literature. The accuracy of dynamics simulation may necessarily needs a more detailed topology for such study. \\

Additionally, the steady-state analysis may not be able to capture islanding issue when a cascading outage occurs \cite{island1,island2}. In a larger system, the initiating event of switching attacks upon the substations may initiate a breaker tripping that results in multiple islands. This can be a unique situation where power flow verification may not always agree with a diverged outcome if multiple islands are discovered and the handling of multiple ``slack'' buses may not immediately compensate the generation-load imbalance across the islands. Without a methodological approach, there may not be a direct way to conclude that stabilities can be achieved among these multiple islands as a result of a cyberattack initially.

}

{\ct
\section{Impact Simulation of Switching Attacks}
The previous section of intelligent attack scenarios on substations is representative of cyberattack impacts of (1) transformer overloading, (2) disruptive switching action of circuit breakers and isolators, and (3) entire substation outages, by attackers who have successfully penetrated into substation control networks. The first two cyberattack scenarios can be visible to dispatchers at control centers before damage has occurred, but only if those personnel are still able to receive alarms from substation equipment while the attack is in progress. These attacks have the long-term impacts to system operation.

We make the assumptions here for impact simulations where (1) there may be more than a single attacker successfully penetrating substation networks, (2) there may be more than one substation network compromised, (3) compromised substation networks may be coordinated among attackers to optimally maximize their attack strategies, for example causing significant impacts from initiating events that can aggravate operating conditions to cascading failure. (4) Attackers have access only to real-time information based on substation networks they compromised – they do not have complete information.

This simulation section is divided into two: (1) steady-state enumeration, (2) dynamic verification. Both simulation approaches have been studied using the IEEE 118-bus system.

\subsection{Steady-State Impact Enumeration}

A steady-state enumeration is an unordered combination of studies by hypothesizing components/substations outages. Such outages do not consider the sequence of switching actions because steady-state enumeration is a pre-screening approach to determine if a certain number of substation outages can lead to divergence of power flow solutions. This is an indication of system collapse that would require a detailed investigation under certain operating conditions of a power grid. Recent studies have shown that certain pivotal substations across an interconnection would have an extreme impact on grid operations, very likely leading to a system-wide cascading blackout. This is one combination of potential problems out of many combinations. The challenge here is computational method to systematically eliminate the worst case scenarios, as a result of reducing the number of IP-based substations plausibility for the purpose of operational planning.

The steady-state enumeration is an unordered combination of study by hypothesizing the components/substations outage. Such outage does not consider the sequence of switching actions because this is a pre-screening approach to determine if a certain number of substations outage can lead to the divergence of power flow solutions. This is an indication of system collapse that would require a detailed investigation under certain operating condition of a power grid. The recent studies have shown that certain pivotal substations across an interconnection would have an extreme impact of the grid operation that is very likely leading to a system-wide cascading blackout \cite{fercstudy,FERCWarns}. This is one of the combinations of potential problems out of many other combinations. The challenge here is computational method to systematically eliminate the worst case scenarios, as a result of reducing the number of IP-based substations plausibility for the purpose of operational planning.

\subsection{Dynamic Impact Verification}
This section provides an intuition of dynamic impact ver-ification from stead-state impact enumerations. While the reverse pyramid model (RPM) \cite{Cybercontingency, cyberrisk, Impact} significantly reduces the number of combinations to consider, substation outage combinations may not be considered critical without dynamic simulation as to the impact of such outages. Part of the reason is that steady-state analysis is based only on power-flow convergence solutions, which may not suggest that a combination of outages can lead to system instability. Another reason dynamic simulation is important is that the steady-state approach does not also provide an evaluation of sequential switching events, and so models any cyberattack on particular substations as one “simultaneous switching action” for all switches. As discussed in Fig. \ref{drawing3}, enumerations can be contradictory, because steady-state simulation screening can be critical where dynamic simulation verification would say otherwise, or vice versa. They both serve as independent evaluators to ensure the consistency of potential cyberattack impact estimates. For example, some combinations can be critical-AND-critical or non-critical-AND-non-critical for steady-state and dynamic analyses, respectively.

Specifically, two possible discrepancies may exist: 1) The power flow converges with a solution but some permutations of dynamic analysis exhibits growing oscillation which lead to no power flow solution in the end of dynamic study, 2) The power flow diverges without a solution but dynamic simulation verification can obtain a post-disturbance operating point. The former example is illustrated in this Section using IEEE 118 bus system shown in Fig. \ref{drawing114}. The elements of this system are with 19 generators, 35 synchronous condensers, 177 lines, 9 transformers, and 91 loads.

\subsubsection{Case 1 – Hypothesized Cyberattack Resulting Dynamical System Instability}
The steady-state impact enumeration has been studied and a single combination, which was deemed non-critical (substations 13, 14, 17, 21, 34), has been selected for dynamic impact verification. This is illustrated in Fig. \ref{drawing114}, which shows the outage of multiple substations as well as the associated transmission circuits connecting with other substations that have been electrically disconnected. 14 line circuits have been disconnected from the system: lines 17–113, 34–43, 34–37, 19–34, 17–31, 21–22, 20–21, 17–18, 16–17, 15–17, 14–15, 12–14, 13–15, and 11-13. Upon execution of the switching action illustrated in purple, the system becomes unstable, and so 11–13 did not open. Under this study, the aforementioned attack switching sequence is executed with an interval of 5 seconds between switching actions. This is under the assumption that attack agents coordinate among the compromised substations. This sequence is an ordered enumeration out of the $14! = 8.718 \times 10 ^ {10}$ ordered combinations. Future research is needed to determine the critical switching action(s) that are dominant to all ordered combinations.
\begin{figure*}
\centering
\includegraphics[width = 18cm]{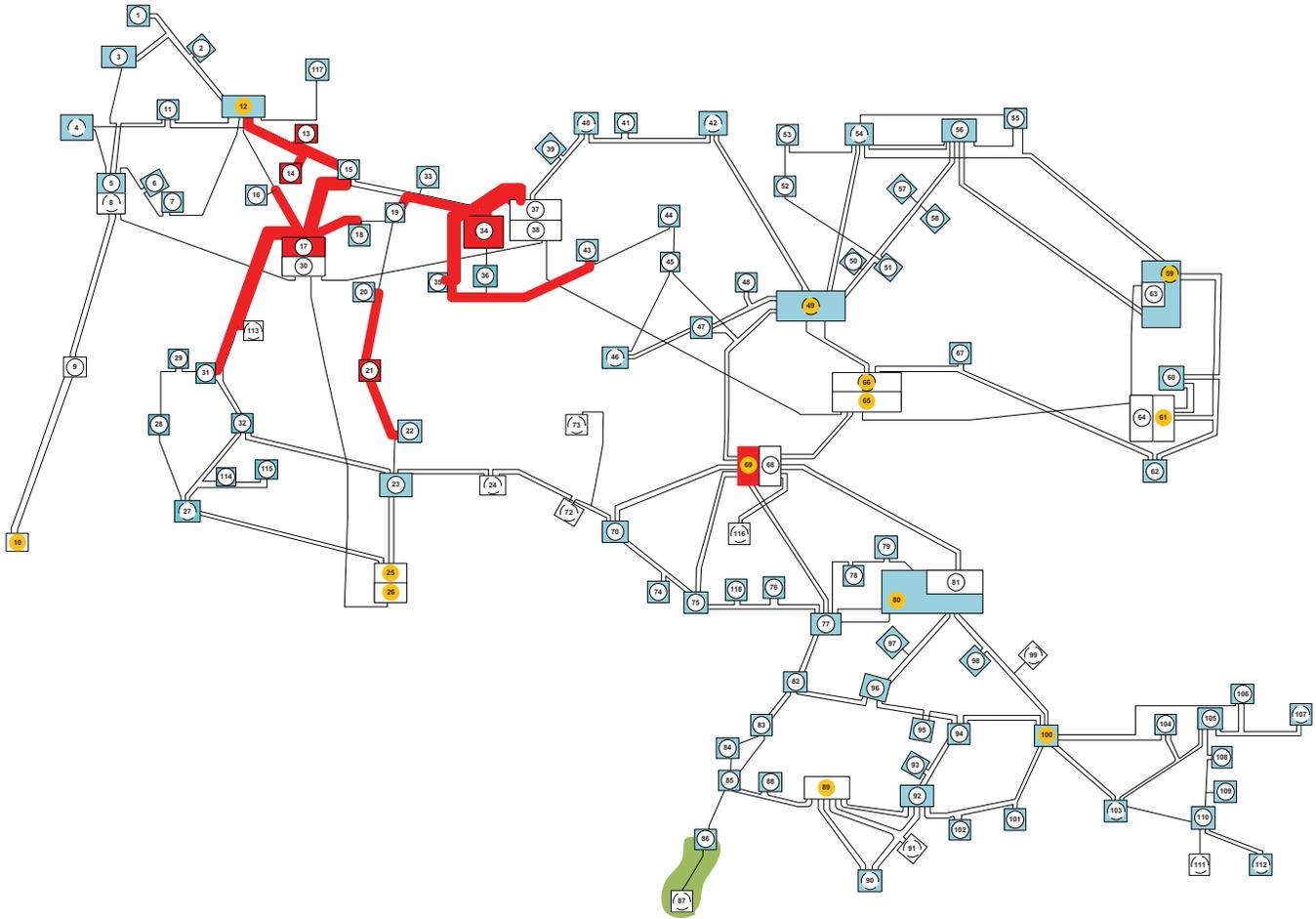}
  \caption{IEEE 118-bus system with Focus of Substations 13, 14, 17, 21, 34 Under Cyberattack.}\label{drawing114} \vspace*{-5pt}
\end{figure*}

Under this permutation, a power flow solution is obtained where outages of substations 13, 14, 17, 21, and 34 are hypothesized. However, this snapshot of study indicates that there is a significantly low voltage in substation 33, which may require corrective voltage control action within a permissible voltage range, typically between 0.94 p.u. and 1.06 p.u. In short, this type of simulation results may show power flow solutions, but with more than one voltage violation or overloaded circuit.

\begin{figure}
\centering
\includegraphics[width = 10cm]{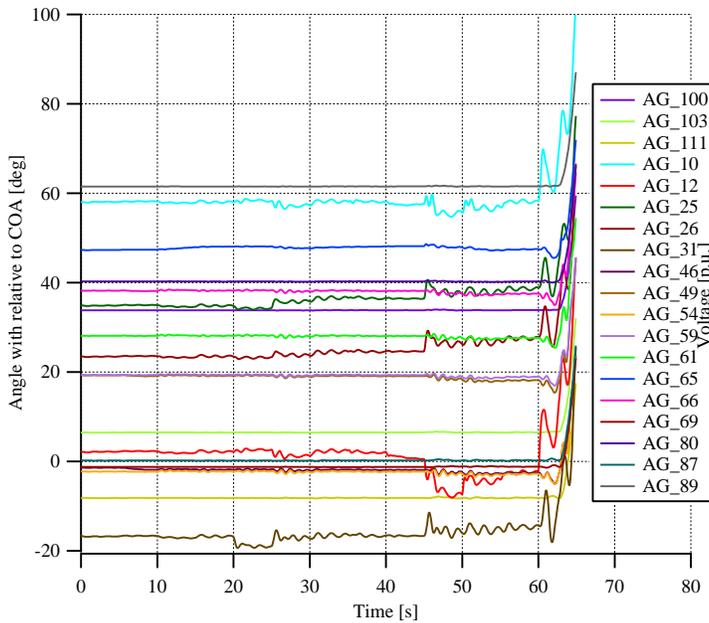}
  \caption{Rotor Angle for 19 Synchronous Machines with Relative to Center of Inertia/Angle (COI/COA).}\label{drawing112}
\end{figure}
Fig. \ref{drawing112} shows the dynamic response of the rotor angle generators with relative to the center of inertia (COI). All generators are assumed to be equipped with Automatic Voltage Regulator (AVR) and a Turbine-Governor Controller, while all synchronous compensators are assumed to be equipped with the AVR only. Because no generic control parameters are given in the test system, example control parameters of a DC exciter and the thermal turbine governor are used in this study. The inertia of those rotating machines are assumed to be 5 seconds each.

\begin{figure}
\centering
\includegraphics[width = 10cm]{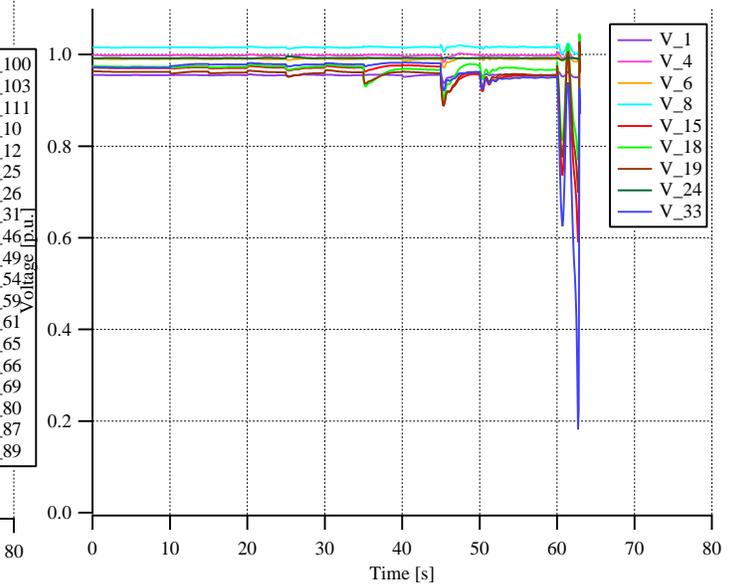}
  \caption{Voltage Response for Substations with Loads in Vicinity.}\label{drawing113}
\end{figure}

It can be clearly observed from Fig. \ref{drawing112} that certain contingencies, such as the removal of line 21-22 at the 25th second, the removal of line 15-17 at the 45-th second, and the removal of line 13-15 at the 60-th second leads to oscillation. All rotor angles of generators show positive damping before 55 seconds under the same the figure. On the other hand, most of the rotor angles of generators increase as growing oscillation occurs after the 13th contingency at 60 seconds. Consequently, the entire system collapses at 63-th second due to this transient instability issue. The last switching action to disconnect line 11-13 was not executed because the system is already extremely vulnerable to collapse. It can also be observed that after the of the synchronous compensator connecting substation 15, 18, and 19 have negative values of relative angles to same COI as shown in Table 3 and therefore, the angle difference between the rotors angle of generators shown in Table 4 and that in Table 3 exceeded 360 degrees at 63 seconds. Therefore, although the power flow solution can be obtained, the system cannot secure the power system stability.

Fig. \ref{drawing113} shows bus voltages. As shown in this figure, the load bus voltage at Bus 33 dramatically decreases after the 13th contingency at 60 seconds. This reveals that the electrical center of the network is located near Bus 33. It can be recognized from Fig \ref{drawing113} that the steady-state analysis based approach cannot take into account dynamic behavior such as power swing oscillation. This means that the steady-state analysis based approach could give optimistic solutions. That means that further research as to how to efficiently search for ``non-critical but close-to critical'' contingency cases will be important.

\subsubsection{Case 2 -- Hypothesized Cyberattack Resulting in Islanding}
This case is the opposite of the case 1 in that there is no critical-case power flow solution, but the dynamic simulation demonstrates a non-critical case. Fig. \ref{drawing200} shows substation 100 that is hypothetically under cyberattack. The unique part of this scenario is that Substation 100 is a pivotal node of the physical system. This means the substation outage would result in two isolated subsystems (Subsystems A and B). The following is the sequential order of attack switching actions:
\begin{enumerate}
  \item line 100--106 at 0-th second
  \item line 100--104 at 5-th second
  \item line 100--103 at 10-th second (two islands are formed)
  \item line 100--101 at 15-th second
  \item line 99--100 at 20-th second
  \item line 98--100 at 25-th second
  \item line 94--100 at 30-th second
  \item line 92--100 at 35-th second
\end{enumerate}

Fig.  \ref{drawing201} consists of two subplots. The above plot is the frequency in subsystem A, which indicates the system is stable. The plot below is the active power plot for the critical generators in subsystem A that shows stabilizing patterns to the new operating points. Subsystem B, however, is unstable in term of frequency prospective as depicted in Fig. \ref{drawing202}. This dynamic simulation does not consider system protection implications. Underfrequency protection would typically shed the loads in subsystem B that may stabilize the subsystem. On the other hand, the underfrequency protection can also electrically disconnect the only generator in subsystem B, leading the subsystem to complete blackout. The outcome here all depends on the rate of change of frequency. Dynamic simulation may not always result in similar observations to stead-state analysis, because the stead-state analysis only includes power flow solutions when there are no other control variables that can be observed on those time-domain simulation patterns. This does not include frequency information in the steady-state analysis. The system frequency in steady-state analysis is always assumed to be a nominal value, such as 60 or 50 Hertz.

\begin{figure*}
\centering
\includegraphics[width = 18cm]{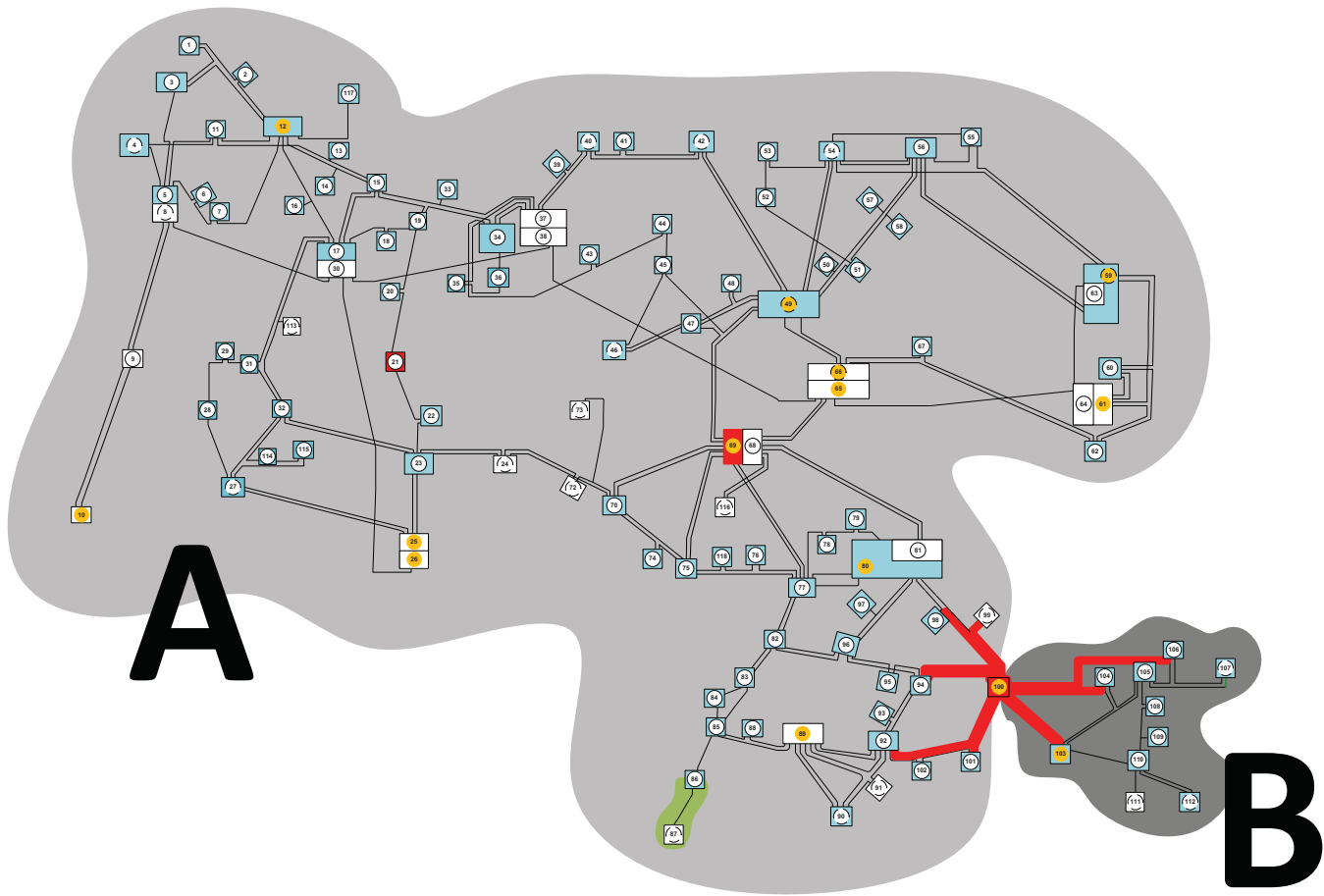}
  \caption{IEEE 118-Bus System with Focus of Substation 100 Under Cyberattack (Form 2 Islands).}\label{drawing200} \vspace*{-5pt}
\end{figure*}

\begin{figure}
\centering
\includegraphics[width = 9cm]{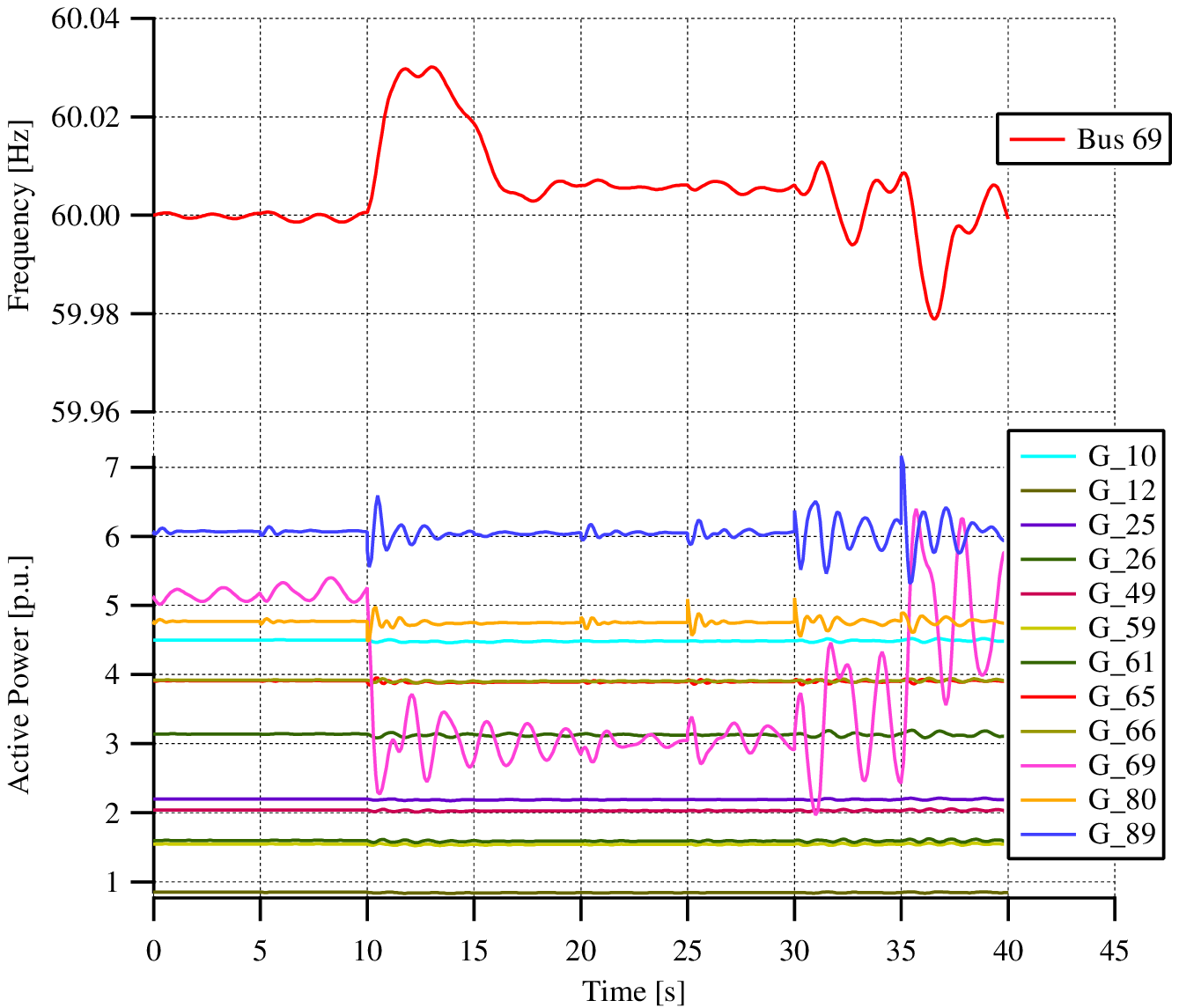}
  \caption{Subsystem A Responses Both in System Frequency and Active Power for Those Generators.}\label{drawing201}
\end{figure}

\begin{figure}
\centering
\includegraphics[width = 9cm]{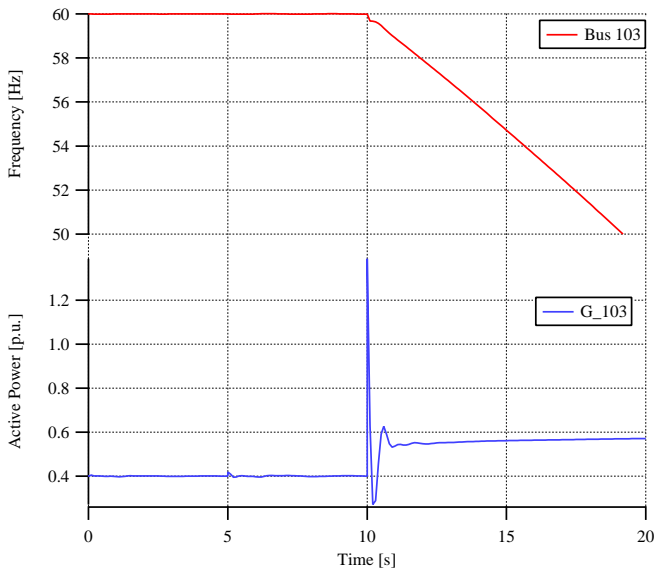}
  \caption{Subsystem B Responses Both in System Frequency and Active Power for the Generator.}\label{drawing202}
\end{figure}
}

\section{Concluding Remarks}
The cybersecurity of a power grid is a topic of research that enables asset owners to anticipate cascading failure as well as identify interdependencies due to cyber-related initiating events. This paper summarizes attack vectors with potential mid-term and long-term effects on the grid. We are in an era where intelligent cyberattackers are emerging with the cross-domain knowledge to execute attack plan against power grids. In such attacks, attackers may not have enough or complete information about the grid to assure the success of their attack plan. Defenders can reduce risks to power infrastructure with improved security analytics in order to anticipate attacks with serious consequences, and thus strategically deploy additional  security protections to critical substations and components. This survey reviews research addressing how determine risk metrics via steady-state and dynamics approaches using existing power system simulators.

{\ct
Improving the cyber-situation awareness of the control centers throughout an interconnection can be challenging. Emerging, renewable energy sources increase the uncertainty of entire power networks, as well as local networks. This uncertainty increases the possibility of unexpected incidents in future networks. System operators will have great difficulty distinguishing  cyberattack incidents from ``normal'' fluctuations related to intermittent renewable generation. Because stability margin could be diminished in future power systems, more than two abnormal power system dynamic phenomena could occur at the same time in the future. The latest research studies have focused on only one abnormal phenomenon. However, the consideration of a number of abnormal phenomena caused by cyberattacks could be strongly required in the future grid. Future work should include implementation of derived metrics on combinations of hypothesized outages to be verified by steady-state and dynamic system simulations. The reduction of permutations and combinations of the hypothesized scenarios can be explored to determine its practicality and systemic bottleneck assessment in order to identify the pivotal components/substations of a power grid.}

\section*{Acknowledgment}
The authors would like to thank R. Bulbul for the preliminary compilation of this work.

\bibliographystyle{IEEEtran}
\bibliography{IEEEabrv,RefDatabase}

\begin{thebibliography}{10}
\providecommand{\url}[1]{#1}
\csname url@samestyle\endcsname
\providecommand{\newblock}{\relax}
\providecommand{\bibinfo}[2]{#2}
\providecommand{\BIBentrySTDinterwordspacing}{\spaceskip=0pt\relax}
\providecommand{\BIBentryALTinterwordstretchfactor}{4}
\providecommand{\BIBentryALTinterwordspacing}{\spaceskip=\fontdimen2\font plus
\BIBentryALTinterwordstretchfactor\fontdimen3\font minus
  \fontdimen4\font\relax}
\providecommand{\BIBforeignlanguage}[2]{{%
\expandafter\ifx\csname l@#1\endcsname\relax
\typeout{** WARNING: IEEEtran.bst: No hyphenation pattern has been}%
\typeout{** loaded for the language `#1'. Using the pattern for}%
\typeout{** the default language instead.}%
\else
\language=\csname l@#1\endcsname
\fi
#2}}
\providecommand{\BIBdecl}{\relax}
\BIBdecl

\bibitem{5484019}
T.~Morris, A.~Srivastava, B.~Reaves, K.~Pavurapu, S.~Abdelwahed, R.~Vaughn,
  W.~McGrew, and Y.~Dandass, ``Engineering future cyber-physical energy
  systems: Challenges, research needs, and roadmap,'' in \emph{North American
  Power Symposium (NAPS)}, 2009, pp. 1--6.

\bibitem{6303885}
S.~Amin, X.~Litrico, S.~Sastry, and A.~Bayen, ``Cyber security of water {SCADA}
  systems - {Part I}: Analysis and experimentation of stealthy deception
  attacks,'' \emph{{IEEE} Trans. Control Syst. Technol.}, vol.~21, no.~5, pp.
  1963--1970, Sep. 2013.

\bibitem{6307833}
------, ``Cyber security of water {SCADA} systems - {Part II}: Attack detection
  using enhanced hydrodynamic models,'' \emph{{IEEE} Trans. Control Syst.
  Technol.}, vol.~21, no.~5, pp. 1679--1693, Sep. 2013.

\bibitem{9000004}
C.-W. Ten, C.-C. Liu, and G.~Manimaran, ``Vulnerability assessment of
  cyberesecurity for {SCADA} systems,'' \emph{{IEEE} Trans. Power Syst.},
  vol.~23, no.~4, pp. 1836--1846, Nov. 2008.

\bibitem{lingfeng2}
Y.~Zhang, Y.~Xiang, L.~Wang, and C.-W. Ten, ``Inclusion of {SCADA} cyber
  vulnerability in power system reliability assessment considering optimal
  resources allocation,'' \emph{{IEEE} Trans. Power Syst.}, vol.~31, no.~6, pp.
  4379--4394, Nov. 2016.

\bibitem{lingfeng1}
Y.~Zhang, Y.~Xiang, and L.~Wang, ``Power system reliability assessment
  incorporating cyberattacks against wind farm energy management systems,''
  \emph{{IEEE} Trans. Smart Grid}, (To appear).

\bibitem{21myth}
L.~Pietre-Cambacedes, M.~Tritschler, and G.~Ericsson, ``Cybersecurity myths on
  power control systems: 21 misconceptions and false beliefs,'' \emph{{IEEE}
  Trans. Power Del.}, vol.~26, no.~1, pp. 161 --172, Jan. 2011.

\bibitem{2}
T.~M. Chen and S.~Abu-Nimeh, ``Lessons from {S}tuxnet,'' \emph{{IEEE} Comput.
  Sci. Eng. Mag.}, vol.~44, no.~4, pp. 91--93, Apr. 2011.

\bibitem{6112333}
F.~Li, A.~Lai, and D.~Ddl, ``{Evidence of Advanced Persistent Threat: A case
  study of malware for political espionage},'' in \emph{proc. 6th Intl. Conf.
  on Malicious and Unwanted Software (MALWARE)}, 2011, pp. 102--109.

\bibitem{3}
{McAfee White Paper}, ``{Global Energy Cyberattacks: Night Dragon},'' Feb.
  2011.

\bibitem{NAE}
D.~W. Cooke, ``The resilience of the electric power delivery system in response
  to terrorism and natural disasters,'' in \emph{Division on Engineering and
  Physical Sciences, Summary of a Workshop, National Research Council of the
  National Academies}, 2013, pp. 1--42.

\bibitem{1IEC61850}
{IEC 61850-6 standard}, \emph{Configuration description language for
  communication in electrical substations related to IEDs}, 1st~ed.\hskip 1em
  plus 0.5em minus 0.4em\relax International Electrotechnical Commision, March,
  2004.

\bibitem{2IEC61850}
L.~Hossenlopp, ``Engineering perspectives on {IEC 61850},'' \emph{{IEEE} Power
  Energy Mag.}, vol.~5, no.~3, pp. 45 -- 50, May 2007.

\bibitem{fercstudy}
\BIBentryALTinterwordspacing
R.~Smith. {U.S.} risks national blackout from small-scale attack. [Online].
  Available:
  \url{http://online.wsj.com/news/articles/SB10001424052702304020104579433670284061220/.}
\BIBentrySTDinterwordspacing

\bibitem{6}
F.~Cleveland, ``{IEC TC57} security standards for the power system's
  information infrastructure - beyond simple encryption,'' in \emph{Proc.
  Trans. and Dist. Conf. and Exhibition, {IEEE-PES}}, May 2006, pp. 1079--1087.

\bibitem{7}
M.~Braendle and R.~Schierholz, ``Cybersecurity for power systems - a closer
  look at the drivers and how to best approach the new challenges,'' in
  \emph{Proc. 64th Annu. Conf. for Protective Relay Engineers}, Apr. 2011, pp.
  322--327.

\bibitem{9ad000004}
C.-W. Ten, G.~Manimaran, and C.-C. Liu, ``Cybersecurity for critical
  infrastructures: attack and defense modeling,'' \emph{{IEEE} Trans. Syst.,
  Man, Cybern. {A}}, vol.~40, no.~4, pp. 853--865, Jul. 2010.

\bibitem{1}
{Government Accountability Office {(GAO)} Report to Congressional Requesters},
  ``Critical infrastructure protection: Multiple efforts to secure control
  systems are under way, but challenges remain,'' vol. GAO-07-1036, Sep. 2007.

\bibitem{RiskAssessment}
J.~Depoy, J.~Phelan, P.~Sholander, B.~Smith, G.~Varnado, and G.~Wyss, ``Risk
  assessment for physical and cyber-attacks on critical infrastructures,''
  \emph{IEEE Military Commun. Conf. (MILCOM)}, Oct. 2005.

\bibitem{8}
S.~Almond, S.~Baird, B.~Flynn, D.~Hawkins, and A.~Mackrell, ``Integrated
  protection and control communications outwith the substation: Cybersecurity
  challenges,'' in \emph{Proc. {IET} 9th Intl. Conf. on Developments in Power
  Syst. Protection {(DPSP)}}, Mar. 2008, pp. 698--701.

\bibitem{10}
M.~Wei and Z.~Chen, ``Reliability analysis of cybersecurity in an electrical
  power system associated {WAN},'' in \emph{Proc. IEEE PES General Meeting},
  Jul. 2012, pp. 1--6.

\bibitem{SimulatedAttack}
\BIBentryALTinterwordspacing
R.~Kuckro. Simulated cyberattack takes down {U.S.} power grid. [Online].
  Available:
  \url{http://www.utilitydive.com/news/simulated-cyberattack-takes-down-us-power-grid/195153/.}
\BIBentrySTDinterwordspacing

\bibitem{NERCstd}
\BIBentryALTinterwordspacing
{NERC Board of Trustees}. (2013, Dec.) Reliability standards for the bulk
  electric systems of north america, {NERC}- standard. [Online]. Available:
  \url{http://www.nerc.com/pa/Stand/Reliability Standards Complete
  Set/RSCompleteSet.pdf.}
\BIBentrySTDinterwordspacing

\bibitem{NERCperform1}
\BIBentryALTinterwordspacing
T.~Burgess, J.~Bian, N.~Khan, and {\emph{et al.}} (2013, May) State of
  reliability 2013 report, {NERC}. [Online]. Available:
  \url{http://www.nerc.com/pa/RAPA/PA/Performance Analysis DL/2013_SOR_May
  15.pdf.}
\BIBentrySTDinterwordspacing

\bibitem{34}
X.~Xu, B.~Lam, R.~Austria, Z.~Ma, Z.~ZHU, R.~Zhu, and J.~Hu, ``Assessing the
  impact of substation-related outages on the network reliability,'' in
  \emph{Proc. Intl. Conf. on Power Syst. Tech. {PowerCon}}, vol.~2, 2002, pp.
  844--848.

\bibitem{35}
X.~Xu, F.~Dong, L.~Huang, and B.~Lam, ``Modeling and simulation of
  substation-related outages in power flow analysis,'' in \emph{Proc. Intl.
  Conf. on Power Syst. Tech. {PowerCon}}, Oct. 2010, pp. 1--5.

\bibitem{32}
K.~Sun and Z.-X. Han, ``Analysis and comparison on several kinds of models of
  cascading failure in power system,'' in \emph{Proc. Trans. and Dist. Conf.
  and Exhibition: Asia and Pacific, IEEE PES}, 2005, pp. 1--7.

\bibitem{33}
L.~Zongxiang, M.~Zhongwei, and Z.~Shuangxi, ``Cascading failure analysis of
  bulk power system using small-world network model,'' in \emph{Proc. Intl.
  Conf. on Probabilistic Methods Applied to Power Syst.}, Sep. 2004, pp.
  635--640.

\bibitem{1664980}
Q.~Chen, C.~Jiang, W.~Qiu, and J.~McCalley, ``Probability models for estimating
  the probabilities of cascading outages in high-voltage transmission
  network,'' \emph{{IEEE} Trans. Power Syst.}, vol.~21, no.~3, pp. 1423--1431,
  2006.

\bibitem{648499}
G.~Ejebe, C.~Jing, J.~Waight, V.~Vittal, G.~Pieper, F.~Jamshidian, P.~Hirsch,
  and D.~Sobajic, ``Online dynamic security assessment in an ems,''
  \emph{{IEEE} Comput. Appl. Power}, vol.~11, no.~1, pp. 43--47, Jan. 1998.

\bibitem{PGOC}
A.~J. Wood, B.~F. Wollenberg, and G.~B. Sheble, \emph{Power Generation
  Operation and Control}, 3rd~ed.\hskip 1em plus 0.5em minus 0.4em\relax Wiley,
  2014.

\bibitem{4113452}
G.~Ejebe and B.~Wollenberg, ``Automatic contingency selection,'' \emph{{IEEE}
  Trans. Power App. Syst.}, vol. PAS-98, no.~1, pp. 97--109, Jan. 1979.

\bibitem{141785}
G.~Ejebe, R.~Paliza, and W.~Tinney, ``An adaptive localization method for
  real-time security analysis,'' \emph{{IEEE} Trans. Power Syst.}, vol.~7,
  no.~2, pp. 777--783, May 1992.

\bibitem{43179}
V.~Brandwajn, ``Efficient bounding method for linear contingency analysis,''
  \emph{{IEEE} Trans. Power Syst.}, vol.~3, no.~1, pp. 38--43, Feb. 1988.

\bibitem{MandatorySingleCont}
\BIBentryALTinterwordspacing
{U.S. Federal Energy Regulatory Commission}. (2007, Apr.) Mandatory reliability
  standards for the bulk-power system, order no. 693, 72 fr 16516. [Online].
  Available: \url{http://www.ferc.gov/whats-new/comm-meet/2007/031507/E-13.pdf}
\BIBentrySTDinterwordspacing

\bibitem{4762167}
J.~Hazra and A.~Sinha, ``Identification of catastrophic failures in power
  system using pattern recognition and fuzzy estimation,'' \emph{{IEEE} Trans.
  Power Syst.}, vol.~24, no.~1, pp. 378--387, 2009.

\bibitem{NERCstdTPL003}
\BIBentryALTinterwordspacing
{NERC Board of Trustees}. (2005, Apr.) System performance following loss of two
  or more bes elements, {NERC}- transmission planning standard tpl-003-0.
  [Online]. Available: \url{www.nerc.com/files/tpl-003-0.pdf}
\BIBentrySTDinterwordspacing

\bibitem{1425dfgf578}
Q.~Chen and J.~McCalley, ``Identifying high risk {N-k} contingencies for
  on-line security assessment,'' \emph{{IEEE} Trans. Power Syst.}, vol.~20,
  no.~2, pp. 823--834, May 2005.

\bibitem{MultipleCont}
C.~Davis and T.~Overbye, ``Multiple element contingency screening,''
  \emph{{IEEE} Trans. Power Syst.}, vol.~26, no.~3, pp. 1294--1301, Aug. 2011.

\bibitem{4110642}
T.~Mikolinnas and B.~Wollenberg, ``An advanced contingency selection
  algorithm,'' \emph{{IEEE} Trans. Power App. Syst.}, vol. PAS-100, no.~2, pp.
  608--617, Feb. 1981.

\bibitem{4110804}
G.~Irisarri and A.~Sasson, ``An automatic contingency selection method for
  on-line security analysis,'' \emph{{IEEE} Trans. Power App. Syst.}, vol.
  PAS-100, no.~4, pp. 1838--1844, Apr. 1981.

\bibitem{5519401}
M.~Enns, J.~Quada, and B.~Sacket, ``Fast linear contingency analysis,''
  \emph{{IEEE} Trans. Power App. Syst.}, vol. PAS-101, no.~4, pp. 783--791,
  Apr. 1982.

\bibitem{MinorJournal1985}
B.~Stott, O.~Alsac, and F.~Alvarado, ``Analytical and computational
  improvements in performance index ranking algorithms for networks,''
  \emph{Intl. Journal of Elec. Power \& Energy Syst.}, vol.~7, no.~3, pp.
  154--160, Jul. 1985.

\bibitem{192968}
G.~Ejebe, H.~V. Meeteren, B.~Wollenberg, and H.~P.~V. Meeteren, ``Fast
  contingency screening and evaluation for voltage security analysis,''
  \emph{{IEEE} Trans. Power Syst.}, vol.~3, no.~4, pp. 1582--1590, Nov. 1988.

\bibitem{1425578}
Q.~Chen and D.~McCalley, ``Identifying high risk n-k contingencies for online
  security assessment,'' \emph{{IEEE} Trans. Power Syst.}, vol.~20, no.~2, pp.
  823--834, May 2005.

\bibitem{4162597}
T.~Guler and G.~Gross, ``Detection of island formation and identification of
  causal factors under multiple line outages,'' \emph{{IEEE} Trans. Power
  Syst.}, vol.~22, no.~2, pp. 505--513, May 2007.

\bibitem{CIGREsession2006}
\BIBentryALTinterwordspacing
{CIGRE C2}. {CIGRE} session {Paris} 2006: Workshop on large disturbances.
  [Online]. Available:
  \url{http://c2.cigre.org/content/download/10140/325796/version/1/file/SC+C2+Minutes+of+Meeting+in+Paris060831ID44VER15.pdf}
\BIBentrySTDinterwordspacing

\bibitem{CIGREsession2008}
\BIBentryALTinterwordspacing
------. {CIGRE} session {Paris} 2008: Workshop on large disturbances. [Online].
  Available: \url{http://c2.cigre.org/What-is-SC-C2/Official-SC-C2-Documents}
\BIBentrySTDinterwordspacing

\bibitem{CIGREsession2010}
\BIBentryALTinterwordspacing
------. {CIGRE} session {Paris} 2010: Workshop on large disturbances. [Online].
  Available: \url{http://www.cigre.org/Events/Session/Session-2010}
\BIBentrySTDinterwordspacing

\bibitem{CIGREsession2012}
\BIBentryALTinterwordspacing
------. {CIGRE} session {Paris} 2012: Workshop on large disturbances. [Online].
  Available: \url{http://c2.cigre.org/Publications/Session-Papers}
\BIBentrySTDinterwordspacing

\bibitem{CIGREsession2014}
\BIBentryALTinterwordspacing
{CIGRE C2 and C5}. {CIGRE} session {Paris} 2014: Workshop on large
  disturbances. [Online]. Available:
  \url{http://www.hro-cigre.hr/session_45_reports_workshop_on_large_disturbance}
\BIBentrySTDinterwordspacing

\bibitem{6089026}
S.~Liu, X.~Feng, D.~Kundur, T.~Zourntos, and K.~Butler-Purry, ``Switched system
  models for coordinated cyber-physical attack construction and simulation,''
  in \emph{proc. 1st {IEEE} Intl. Workshop on Smart Grid Modeling and
  Simulation {(SGMS)}}, Oct. 2011, pp. 49--54.

\bibitem{6112807}
M.~Vaiman, K.~Bell, Y.~Chen, B.~Chowdhury, I.~Dobson, P.~Hines, M.~Papic,
  S.~Miller, and P.~Zhang, ``Risk assessment of cascading outages:
  Methodologies and {C}hallenges,'' \emph{{IEEE} Trans. Power Syst.}, vol.~27,
  no.~2, pp. 631--641, May. 2012.

\bibitem{1306711}
D.~Gerbec, S.~Gasperic, I.~Smon, and F.~Gubina, ``Determining the load profiles
  of consumers based on fuzzy logic and probability neural networks,''
  \emph{{IEEE} Trans. Power Syst.}, vol. 151, no.~3, pp. 395--400, May 2004.

\bibitem{Cybercontingency}
R.~Bulbul, C.-W. Ten, and A.~Ginter, ``Cyber-contingency evaluation for
  multiple hypothesized substation outages,'' in \emph{Proc. Innovative Smart
  Grid Technologies Conference (ISGT), IEEE PES}, Feb. 2014, pp. 1--5.

\bibitem{cyberrisk}
------, ``Risk evaluation for hypothesized multiple busbar outages,'' in
  \emph{Proc. IEEE PES General Meeting Conf. and Exposition}, Jul. 2014, pp.
  1--5.

\bibitem{CyberConJournal}
C.-W. Ten, A.~Ginter, and R.~Bulbul, ``Cyber-based contingency analysis,''
  \emph{{IEEE} Trans. Power Syst.}, vol.~31, no.~4, pp. 3040--3050, Jul. 2016.

\bibitem{Impact}
R.~Bulbul, Y.~Gong, C.-W. Ten, A.~Ginter, and S.~Mei, ``Impact quantification
  of hypothesized attack scenarios on bus differential relays,'' in \emph{Proc.
  18th Power Systems Computation Conf.}, Aug. 2014, pp. 1--7.

\bibitem{RealibityImpact}
J.~Stamp, A.~McIntyre, and B.~Ricardson, ``Reliability impacts from cyberattack
  on electric power systems,'' in \emph{Proc. IEEE PES Power Systems Conference
  and Exposition}, Seattle, Washington, 2009, pp. 1--8.

\bibitem{6035612}
M.~Zeller, ``Myth or reality - does the aurora vulnerability pose a risk to my
  generator?'' in \emph{Proc. 64th Annual Conference for Protective Relay
  Engineers}, 2011, pp. 130--136.

\bibitem{algorithm}
R.~Sedgewick, \emph{Algorithms in C++}, 3rd~ed.\hskip 1em plus 0.5em minus
  0.4em\relax Pearson Education, Inc., Dec., 2001.

\bibitem{5325912}
``{IEEE} guide for protective relay applications to power system buses,''
  \emph{IEEE Std C37.234-2009}, pp. 1--115, Nov. 2009.

\bibitem{6938963}
B.~Chen, K.~L. Butler-Purry, S.~Nuthalapati, and D.~Kundur, ``Network delay
  caused by cyber attacks on svc and its impact on transient stability of smart
  grids,'' in \emph{Proc. IEEE PES General Meeting}, Jul. 2014, pp. 1--5.

\bibitem{6175560}
A.~Stefanov and C.-C. Liu, ``Cyber-power system security in a smart grid
  environment,'' in \emph{2012 IEEE 51st Annual Conference on Decision and
  Control}, 2012, pp. 1--3.

\bibitem{6473865}
A.~Hahn, A.~Ashok, S.~Sridhar, and M.~Govindarasu, ``Cyber-physical security
  testbeds: Architecture, application, and evaluation for smart grid,''
  \emph{{IEEE} Trans. Smart Grid}, vol.~4, no.~2, pp. 847--855, Jun. 2013.

\bibitem{AaH}
C.~Wang, C.-W. Ten, Y.~Hou, and A.~Ginter, ``Cyber inference system for
  substation anomalies against alter-and-hide attacks,'' \emph{{IEEE} Trans.
  Power Syst.}, pp. 1--14.

\bibitem{6666849}
B.~Chen, K.~L. Butler-Purry, and D.~Kundur, ``Impact analysis of transient
  stability due to cyber attack on facts devices,'' in \emph{Proc. North
  American Power Symposium (NAPS)}, 2013, pp. 1--6.

\bibitem{6672740}
B.~Chen, S.~Mashayekh, K.~L. Butler-Purry, and D.~Kundur, ``Impact of cyber
  attacks on transient stability of smart grids with voltage support devices,''
  in \emph{Proc. IEEE PES General Meeting}, Jul. 2013, pp. 1--5.

\bibitem{6426269}
P.~M. Esfahani, G.~A. M.~Vrakopoulou, and J.~Lygeros, ``A tractable nonlinear
  fault detection and isolation technique with application to the
  cyber-physical security of power systems,'' in \emph{2012 IEEE 51st Annual
  Conference on Decision and Control}, 2012, pp. 3433--3438.

\bibitem{6740883}
S.~Sridhar and M.~Govindarasu, ``Model-based attack detection and mitigation
  for automatic generation control,'' \emph{{IEEE} Trans. Smart Grid}, vol.~5,
  no.~2, pp. 580--591, Mar. 2015.

\bibitem{6672731}
J.~Yan, M.~Govindarasu, C.-C. Liu, and U.~Vaidya, ``A pmu-based risk assessment
  framework for power control systems,'' in \emph{Proc. IEEE PES General
  Meeting}, 2013, pp. 1--5.

\bibitem{6848210}
M.~Wei and W.~Wang, ``Greenbench: A benchmark for observing power grid
  vulnerability under data-centric threats,'' in \emph{Proc. IEEE Conf. on
  Comp. Comm.}, 2014, pp. 2625--2633.

\bibitem{ZhuHan}
H.~Li and Z.~Han, ``Manipulating the electricity power market via jamming the
  price signaling in smart grid,'' in \emph{Proc. IEEE Intl. Workshop on Smart
  Grid Comm. and Networks}, 2011, pp. 1168--1172.

\bibitem{6965381}
B.~Chen, K.~L. Butler-Purry, A.~Goulart, and D.~Kundur, ``Implementing a
  real-time cyber-physical system test bed in rtds and opnet,'' in \emph{Proc.
  North American Power Symposium (NAPS)}, 2014, pp. 1--6.

\bibitem{6062375}
Y.~Susuki, T.~J. Koo, H.~Ebina, T.~Yamazaki, T.~Ochi, T.~Uemura, and
  T.~Hikihara, ``A hybrid system approach to the analysis and design of power
  grid dynamic performance,'' \emph{Proc. {IEEE}}, vol. 100, no.~1, pp.
  225--239, Jan. 2012.

\bibitem{6344956}
R.~M. Kolacinski and K.~A. Loparo, ``A mathematic framework for analysis of
  complex cyber-physical power systems,'' \emph{{IEEE} Trans. Smart Grid},
  vol.~3, no.~3, pp. 1444--1456, Sep. 2012.

\bibitem{23}
T.~Weekes, T.~Molinski, and G.~Swift, ``Transient transformer overload ratings
  and protection,'' \emph{{IEEE} Electr. Insul. Mag.}, vol.~20, no.~2, pp.
  32--35, Mar.-Apr. 2004.

\bibitem{24}
P.~Sen and S.~Pansuwan, ``Overloading and loss-of-life assessment guidelines of
  oil-cooled transformers,'' in \emph{Proc. Rural Electric Power Conf.}, Apr. -
  May 2001, pp. B4/1--B4/8.

\bibitem{26}
{The Institute of Electrical and Electronics Engineers, Inc.}, ``{IEEE} guide
  for loading mineral-oil-immersed transformers,'' \emph{IEEE Std C57.91-1995},
  pp. 1--112, 1996.

\bibitem{25}
J.~Perez, ``Fundamental principles of transformer thermal loading and
  protection,'' in \emph{Proc. 63rd Annu. Conf. for Protective Relay
  Engineers}, Apr. 29 2010, pp. 1--14.

\bibitem{28}
S.~Kharin, ``Role of metallic vapor pressure in contact bouncing and welding at
  closure of electrical contacts in vacuum,'' in \emph{Proc. 58th {IEEE} Conf.
  on Electrical Contacts {(Holm)}}, Sep. 2012, pp. 1--7.

\bibitem{30}
J.~Barker, W.~Beran, K.~Hendrix, M.~Hill, P.~Kolarik, and D.~Massey,
  ``Determination of disconnecting switch ratings for the {Pennsylvania-New
  Jersey-Maryland} interconnection,'' \emph{{IEEE} Trans. Power App. Syst.},
  vol. PAS-91, no.~2, pp. 404 --411, Mar. 1972.

\bibitem{27}
T.~Lindquist, L.~Bertling, and R.~Eriksson, ``Circuit breaker failure data and
  reliability modelling,'' \emph{{IET} Gener. Trans. Dist.}, vol.~2, no.~6, pp.
  813--820, Nov. 2008.

\bibitem{cascading1}
\BIBentryALTinterwordspacing
M.~Rios, K.~Bell, D.~Kirschen, and R.~Allan. (1999) Computation of the value of
  security. Mannchester Centre for Electrical Energy, UMIST. [Online].
  Available:
  \url{http://www2.ee.washington.edu/research/real/Library/Reports/Value_of_Security_Part_I.pdf}
\BIBentrySTDinterwordspacing

\bibitem{cascading2}
\BIBentryALTinterwordspacing
P.~Rezaei. (2015) Cascading failure risk estimation and mitigation in power
  systems. University of Vermont. [Online]. Available:
  \url{http://scholarworks.uvm.edu/cgi/viewcontent.cgi?article=1481\&context=\\graddis}
\BIBentrySTDinterwordspacing

\bibitem{cascading3}
J.~Yan, Y.~Tang, H.~He, and Y.~Sun, ``Cascading failure analysis with {DC}
  power flow model and transient stability analysis,'' \emph{{IEEE} Trans.
  Power Syst.}, vol.~30, no.~1, pp. 285--297, Jan. 2015.

\bibitem{cascading4}
M.~J. Eppstein and P.~D.~H. Hines, ``A ``random chemistry'' algorithm for
  identifying collections of multiple contingencies that initiate cascading
  failure,'' \emph{{IEEE} Trans. Power Syst.}, vol.~27, no.~3, pp. 1698--1705,
  Aug. 2012.

\bibitem{island1}
R.~Sun, Z.~Wu, and V.~A. Centeno, ``Power system islanding detection \&
  identification using topology approach and decision tree,'' in \emph{Power
  and Energy Society General Meeting, 2011 IEEE}, San Diego, CA, July 2011, pp.
  1--6.

\bibitem{island2}
L.~Ding, F.~M. Gonzalez-Longatt, P.~Wall, and V.~Terzija, ``Two-step spectral
  clustering controlled islanding algorithm,'' \emph{IEEE Trans. Power Syst.},
  vol.~28, no.~1, pp. 75--84, Feb. 2013.

\bibitem{FERCWarns}
\BIBentryALTinterwordspacing
K.~Jorgustin, ``Nine electric power grid substations will bring it all down,''
  Mar. 17 2014. [Online]. Available:
  \url{http://modernsurvivalblog.com/systemic-risk/9-electric-power-grid-substations-will-bring-it-all-down/}
\BIBentrySTDinterwordspacing

\end{thebibliography}

\begin{IEEEbiography}[{\includegraphics[width=1in,height=1.25in,clip,keepaspectratio]{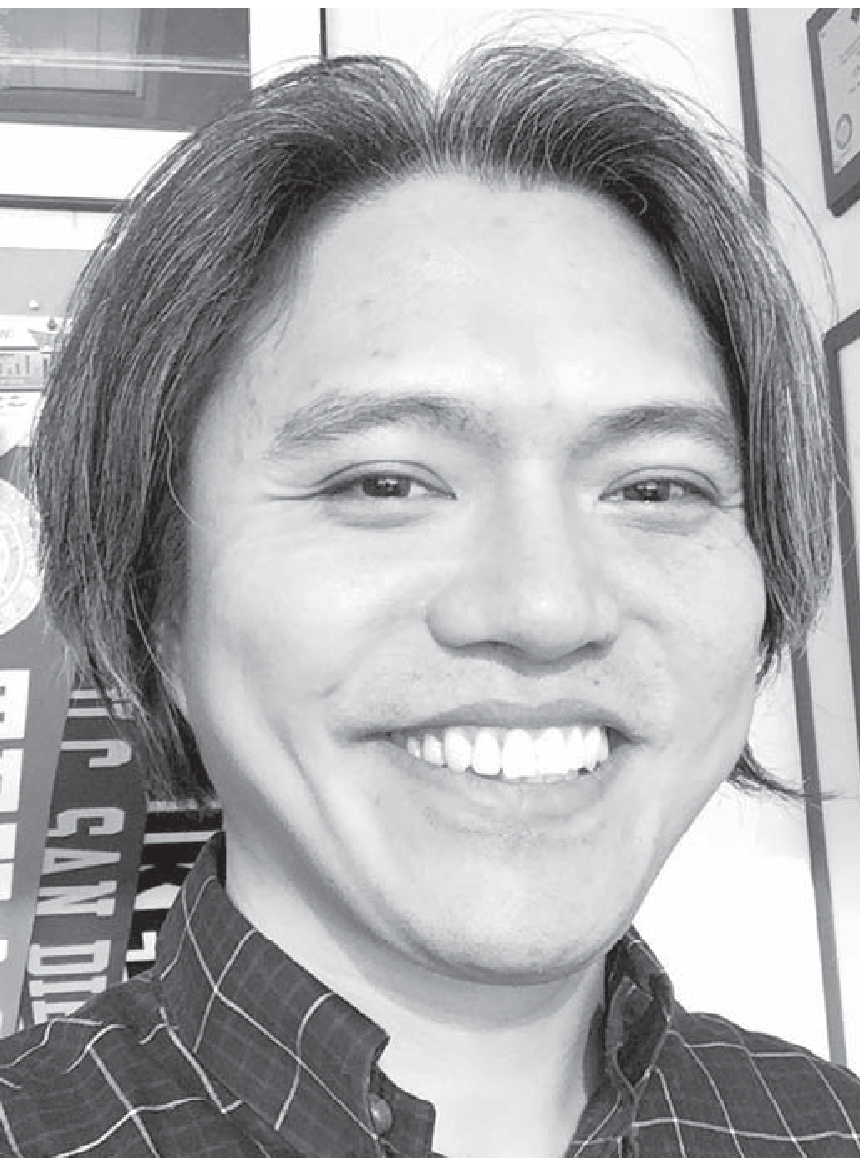}}]
{Chee-Wooi Ten} received the B.S. and M.S. degrees in Electrical Engineering from Iowa State University, Ames, in 1999 and 2001, respectively. He was an Application Engineer with Siemens Energy Management and Information System (SEMIS) in Singapore from 2002 to 2006. He later received the Ph.D. degree in 2009 from University College Dublin (UCD).

He is currently an Associate Professor of Electrical and Computer Engineering at Michigan Technological University. His primary research interests are SCADA cybersecurity, modeling of interdependencies for critical cyberinfrastructures, and power automation applications. Dr. Ten serves as an editor for IEEE Transactions on Smart Grid and Elsevier Journal Sustainable Energy, Grids and Networks (SEGAN).
\end{IEEEbiography}

\begin{IEEEbiography}[{\includegraphics[width=1in,height=1.25in,clip,keepaspectratio]{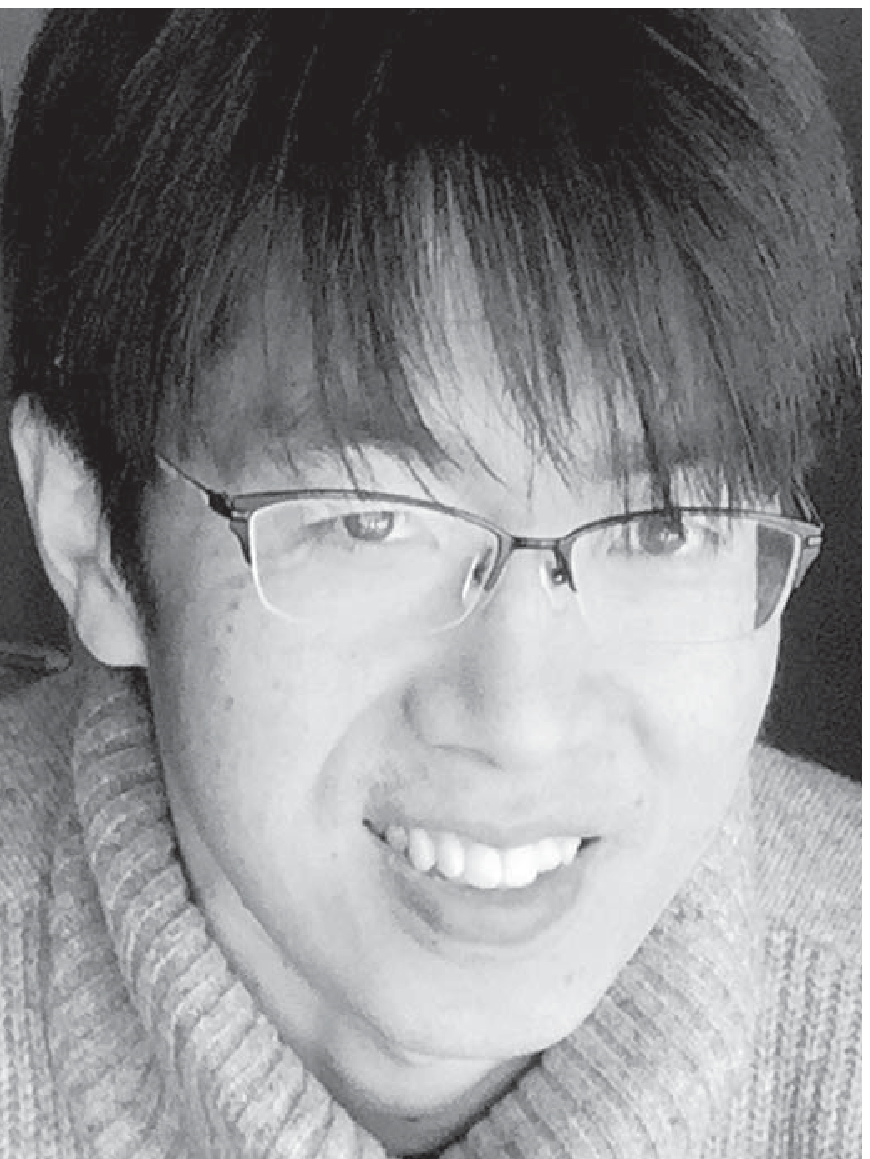}}]
{Koji Yamashita} received his B.S. and M.S. degrees in Electrical Engineering from Waseda University, Tokyo, Japan, in 1993 and 1995, respectively.  Mr. Yamashita was a visiting researcher at Iowa State University during 2006-2007. He is a member of IEEE and the Institute of Electrical Engineers of Japan (IEEJ).

He is currently a Senior Research Scientist at the Central Research Institute of Electric Power Industry (CRIEPI) in Tokyo, Japan and has been with the Department of Power Systems since 1995. He is also working towards his doctoral degree at Michigan Technological University. His areas of interest include hypothesized attack scenarios and its resulting impact on system dynamics and stability, wide-area protection and control as well as generation/load imbalance.
\end{IEEEbiography}

\begin{IEEEbiography}[{\includegraphics[width=1in,height=1.25in,clip,keepaspectratio]{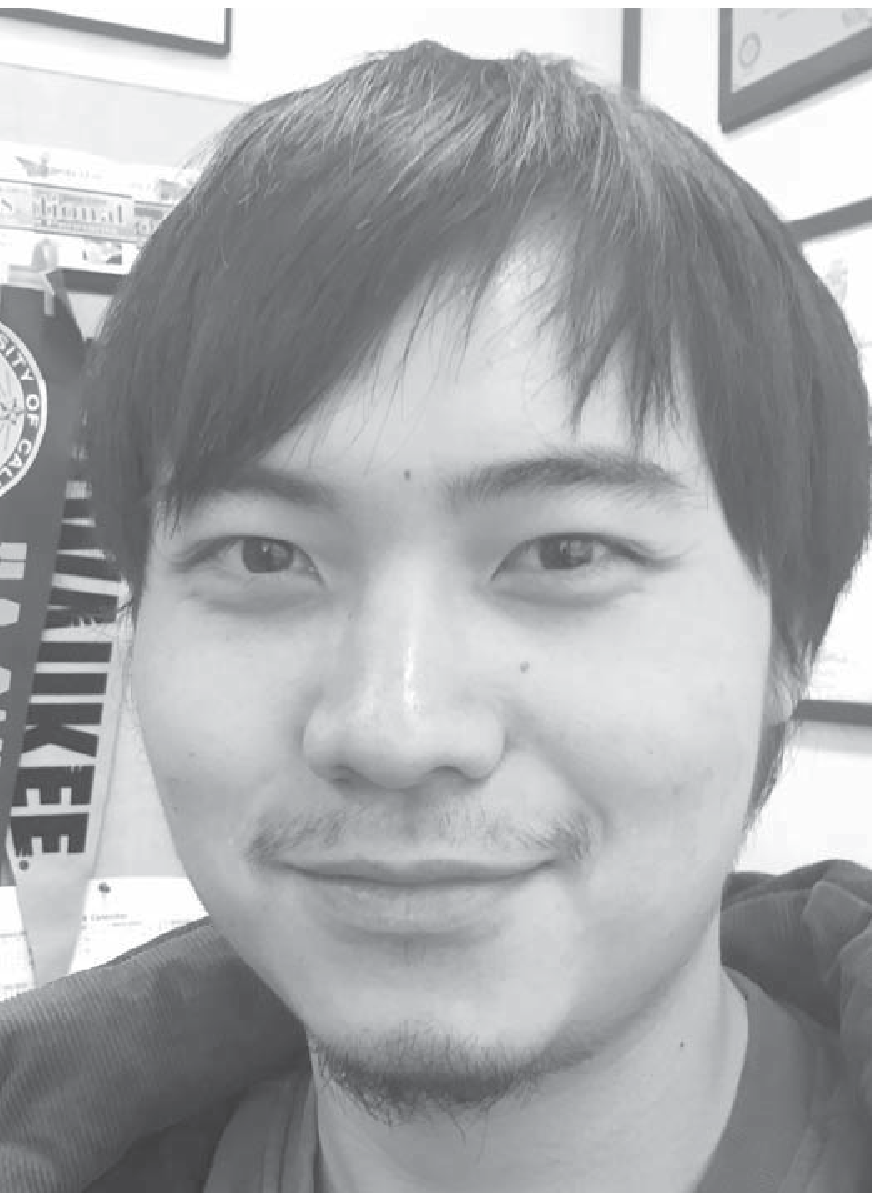}}]
{Zhiyuan Yang} received his B.S. degree in Electrical Engineering and Automation from North China Electric Power University in 2015. He was working as a student member in electronics laboratory and microcomputer relay protection laboratory at North China Electric Power University (NCEPU).

He is currently pursuing a PhD student in the Electrical and Computer Engineering Department at Michigan Technological University. His research interests include combinatorics, and electronics circuit. His areas of interest include combinatorial evaluation of hypothesized attack scenarios and effective enumeration of systemic impact using steady-state approach.
\end{IEEEbiography}

\begin{IEEEbiography}[{\includegraphics[width=1in,height=1.25in,clip,keepaspectratio]{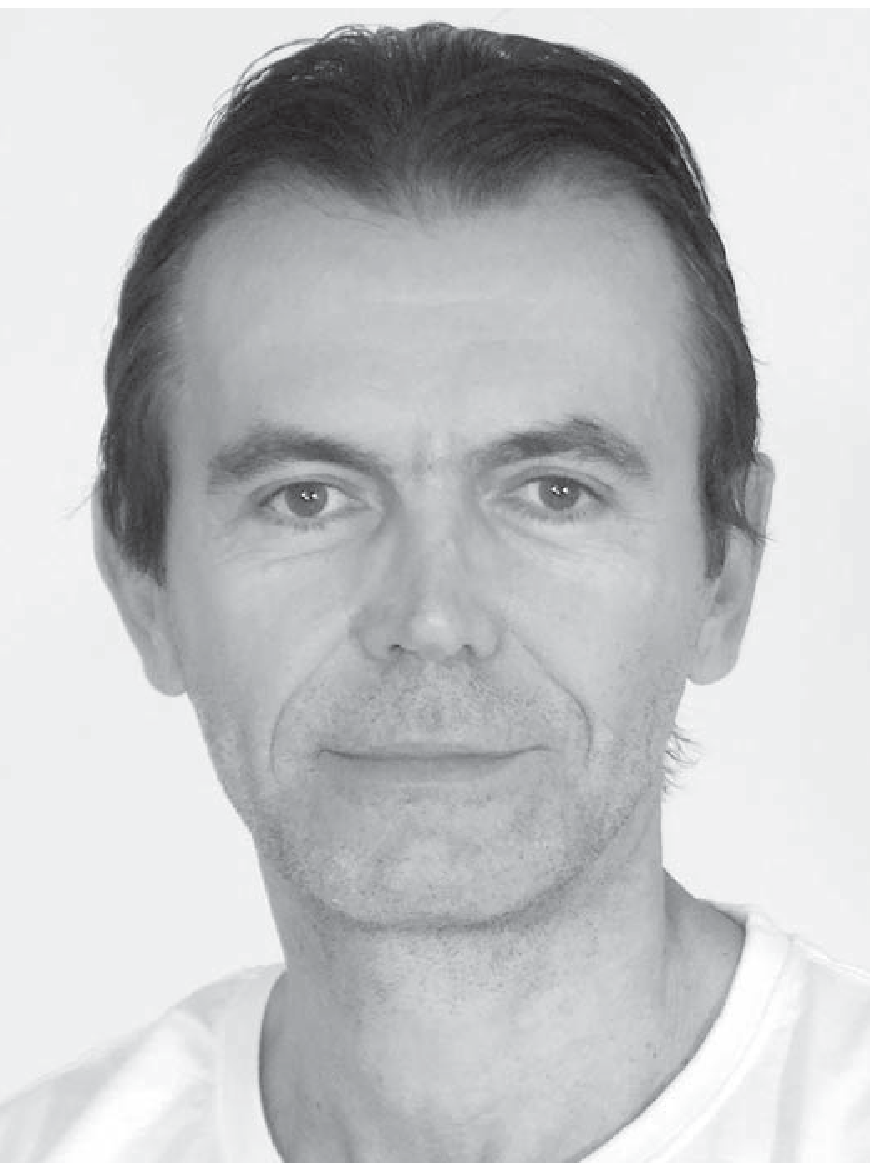}}]
{Athanasios V. Vasilakos} is a Visiting Professor with the Lule{\aa} University of Technology, Sweden. He served or is serving as an Editor for many technical journals, such as the IEEE Transactions on Network and Service Management; IEEE Transactions on Cloud Computing, IEEE Transactions on Information Forensics and Security, IEEE Transactions on Cybernetics; IEEE Transactions on Nanobioscience; IEEE Transactions on Information Technology in Biomedicine; ACM Transactions on Autonomous and Adaptive Systems; the IEEE Journal on Selected Areas in Communications. He is also General Chair of the European Alliances for Innovation (www.eai.eu).
\end{IEEEbiography}

\begin{IEEEbiography}[{\includegraphics[width=1in,height=1.25in,clip,keepaspectratio]{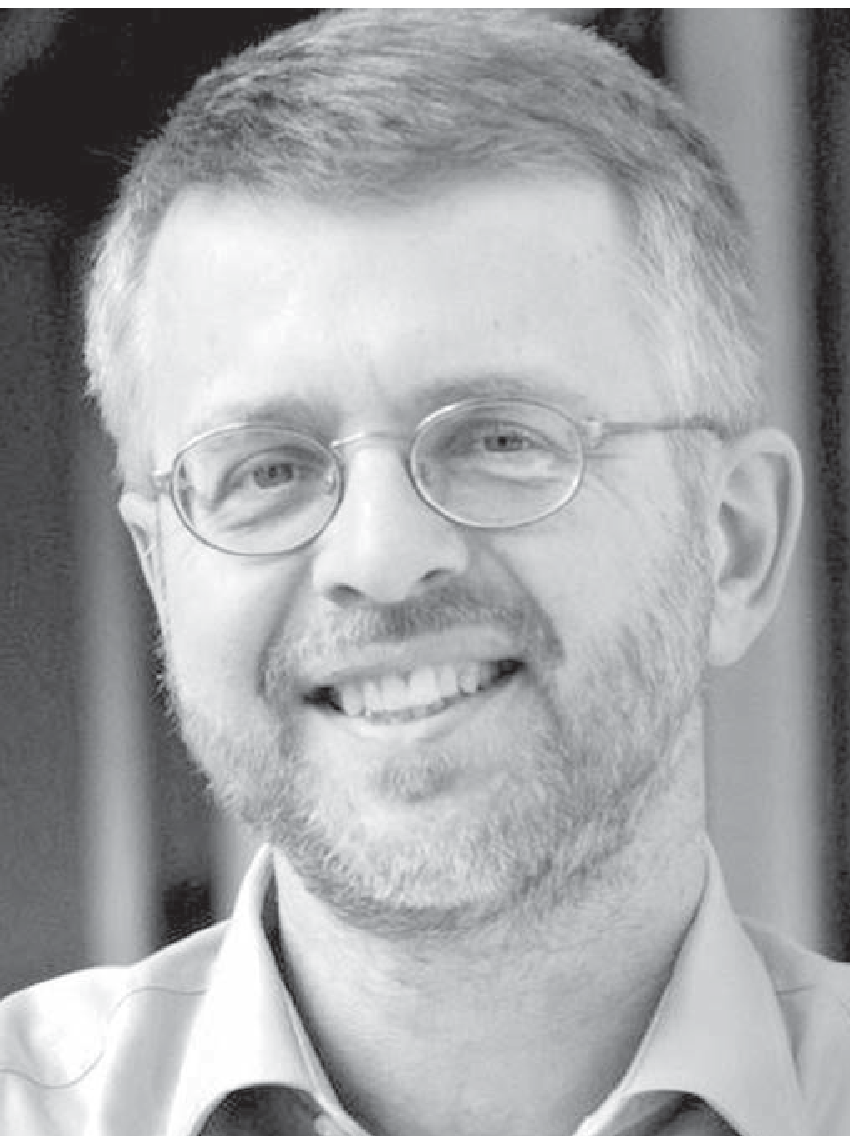}}]
{Andrew Ginter} is the Vice President of Industrial Security at Waterfall Security Solutions. He spent the first part of his career developing systems level and control system products for a number of vendors, including Honeywell and Hewlett-Packard. At Agilent Technologies, he led development of middleware products connecting industrial control systems to the SAP enterprise resource planning systems. As Chief Technology Officer at Industrial Defender, Andrew led the development of hte core industrial security product suite.

At present, Andrew represents Waterfall Security Solutions on standards bodies and works with customers to incorporate Waterfall Unidirectional Gateways into their induial network designs. Andrew Holds degree in Mathematics and Computer Science from the University of Calgary, as well as Industrial Security Professional (ISP), Information Technology Certified Professional (ITCP), and Certified Information Systems Security Professional (CISSP) accreditations.
\end{IEEEbiography}

\end{document}